# Effective Capacity of Two-Hop Wireless Communication Systems

Deli Qiao, Mustafa Cenk Gursoy, and Senem Velipasalar

## Abstract

A two-hop wireless communication link in which a source sends data to a destination with the aid of an intermediate relay node is studied. It is assumed that there is no direct link between the source and the destination, and the relay forwards the information to the destination by employing the decode-and-forward scheme. Both the source and intermediate relay nodes are assumed to operate under statistical quality of service (QoS) constraints imposed as limitations on the buffer overflow probabilities. The maximum constant arrival rates that can be supported by this two-hop link in the presence of QoS constraints are characterized by determining the effective capacity of such links as a function of the QoS parameters and signal-to-noise ratios at the source and relay, and the fading distributions of the links. The analysis is performed for both full-duplex and half-duplex relaying. Through this study, the impact upon the throughput of having buffer constraints at the source and intermediate relay nodes is identified. The interactions between the buffer constraints in different nodes and how they affect the performance are studied. The optimal time-sharing parameter in half-duplex relaying is determined, and performance with half-duplex relaying is investigated.

## I. INTRODUCTION

Fueled by the fourth generation (4G) wireless standards, smart phones and tablets, social networking tools and video-sharing sites, wireless transmission of multimedia content has significantly increased in

The authors are with the Department of Electrical Engineering, University of Nebraska-Lincoln, Lincoln, NE 68588 (e-mails: dqiao726@huskers.unl.edu, gursoy@engr.unl.edu, velipasa@engr.unl.edu).

This work was supported by the National Science Foundation under Grants CNS–0834753, and CCF–0917265.

volume and is expected to be the dominant traffic in data communications. Such wireless multimedia traffic requires certain quality-of-service (QoS) guarantees so that acceptable performance and quality levels can be met for the end-users. For instance, in voice over IP (VoIP), interactive-video (e.g., videoconferencing), and streaming-video applications in wireless systems, latency is a key QoS metric. In such cases, information has to be communicated with minimal delay. Hence, certain constraints on the queue length should be imposed in order to have the data not wait too long in the buffer at the transmitter. At the same time, satisfying these QoS considerations is challenging in wireless communication scenarios. Due to mobility, changing environment and multipath fading, the power of the received signal, and hence the instantaneous rates supported by the channel, fluctuate randomly [1]. In such a volatile environment, providing deterministic delay guarantees either is not possible or, when it is possible, requires the system to operate pessimistically and achieve low performance underutilizing the resources. Therefore, wireless systems are better suited to support statistical QoS guarantees.

In [2], Chang employed the effective bandwidth theory to analyze systems operating under statistical QoS constraints. These constraints are imposed on buffer violation probabilities and are specified by the QoS exponent $\theta$, which is defined as

$$\lim_{Q_{\max} \to \infty} \frac{\log \Pr\{Q > Q_{\max}\}}{Q_{\max}} = -\theta, \tag{1}$$

where $Q$ is the queue length in steady state and $Q_{\max}$ is a threshold indicating the maximal tolerable queue length. If the above limiting formulation is satisfied, then the buffer violation probability behaves as $\Pr\{Q > Q_{\max}\} \approx e^{-\theta Q_{\max}}$ for large $Q_{\max}$. Therefore, QoS exponent $\theta$ is the exponential decay rate of the buffer overflow probability for large $Q_{\max}$. A larger $\theta$ implies a lower probability of violating the queue length and is a more stringent QoS constraint. In [3], Chang and Zajic characterized the effective bandwidths of the time varying departure processes. In [4], Chang and Thomas applied the effective bandwidth theory to high-speed digital networks. More recently, Wu and Negi in [5] defined the dual concept of effective capacity, which provides the maximum constant arrival rate that can be supported by a given departure or service process while satisfying statistical QoS constraints. The analysis and application of effective capacity in various settings have attracted much interest recently (see e.g. [6]-[13] and references therein). For instance, optimal power control policies that maximize the effective capacity of a point-to-point link have been derived in [6]. In [10], the authors study the multiple-input single-output (MISO) channels and determine the optimal transmit strategies with covariance feedback when effective capacity is adopted as the performance metric. In [11], effective capacity in a time-division-based downlink system is characterized, and optimal scheduling schemes that achieve the points on the boundary of the effective capacity region are identified.

In this paper, we consider two-hop wireless links and investigate the throughput in the presence of QoS constraints by studying the effective capacity. We note that references [12] and [13] have also recently investigated the effective capacity of relay channels. Tang and Zhang in [12] analyzed the power allocation policies in relay networks under the assumption that the relay node has no buffer constraints. Parag and Chamberland in [13] provided a queueing analysis of a butterfly network with constant rate for each link. However, they assumed that there is no congestion at the intermediate nodes. In this work, as a significant departure from previous studies, we assume that both the source and the relay nodes are subject to QoS constraints specified by the QoS exponents $\theta_1$ and $\theta_2$. Now, we face a more challenging scenario in which the buffer constraints at the source and relay interact. Moreover, we consider a general relay channel model in which the fading coefficients for each link can have arbitrary distributions. We concentrate on the decode-and-forward (DF) relaying scheme. Assuming that the relay operates in full-duplex or half-duplex mode, we determine the effective capacity as a function of $\theta_1$ and $\theta_2$. Through this analysis, we characterize the impact of the presence of QoS constraints at the relay and also of half-duplex operation on the throughput of the two-hop link.

The rest of this paper is organized as follows. In Section II, the system model and necessary preliminaries are provided. In Section III, we describe our main results on the effective capacity and present numerical results. Finally, in Section IV, we conclude the paper. Lengthy proofs are relegated to the Appendix.

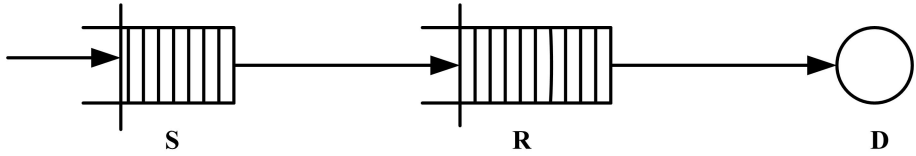

Fig. 1. The system model.

## II. SYSTEM MODEL AND PRELIMINARIES

### A. System Model

The two-hop communication link is depicted in Figure 1. In this model, source **S** is sending information to the destination **D** with the help of the intermediate relay node **R**. We assume that there is no direct link between **S** and **D** (which, for instance, holds, if these nodes are sufficiently far apart in distance). Both the source and the intermediate relay node operate under QoS constraints (i.e., buffer constraints) specified by the QoS exponents $\theta_1$ and $\theta_2$, respectively. Hence, the source and relay buffer violation probabilities should, for some large $Q_{\max}$, satisfy

$$\Pr\{Q_s \geq Q_{\max}\} \approx e^{-\theta_1 Q_{\max}} \tag{2}$$

and

$$\Pr\{Q_r \geq Q_{\max}\} \approx e^{-\theta_2 Q_{\max}}, \tag{3}$$

respectively. Above, $Q_s$ and $Q_r$ denote the stationary queue lengths at the source and relay, respectively.

We consider both full-duplex and half-duplex relay operation. The full-duplex relay can receive and transmit simultaneously while the half-duplex relay first listens and then transmits. Therefore, reception and transmission at the half-duplex relay occur in non-overlapping intervals.

Next, we identify the discrete-time input and output relationships. In the $i^{\text{th}}$ symbol duration, the signal $Y_r$ received at the relay from the source and the signal $Y_d$ received at the destination from the relay can be

expressed as

$$Y_r[i] = g_1[i]X_1[i] + n_1[i] \qquad (4)$$

$$Y_d[i] = g_2[i]X_2[i] + n_2[i] \qquad (5)$$

where $X_j$ for $j = \{1, 2\}$ denote the inputs for the links $\mathbf{S}-\mathbf{R}$ and $\mathbf{R}-\mathbf{D}$, respectively. More specifically, $X_1$ is the signal sent from the source and $X_2$ is sent from the relay. The inputs are subject to individual average energy constraints $\mathbb{E}\{|X_j|^2\} \leq \bar{P}_j/B, j = \{1, 2\}$ where $B$ is the bandwidth. Assuming that the symbol rate is $B$ complex symbols per second, we can easily see that the symbol energy constraint of $\bar{P}_j/B$ implies that the channel input has a power constraint of $\bar{P}_j$. We assume that the fading coefficients $g_j, j = \{1, 2\}$ are jointly stationary and ergodic discrete-time processes, and we denote the magnitude-square of the fading coefficients by $z_j[i] = |g_j[i]|^2$. Above, in the channel input-output relationships, the noise component $n_j[i]$ is a zero-mean, circularly symmetric, complex Gaussian random variable with variance $\mathbb{E}\{|n_j[i]|^2\} = N_j$ for $j = 1, 2$. The additive Gaussian noise samples $\{n_j[i]\}$ are assumed to form an independent and identically distributed (i.i.d.) sequence. We denote the signal-to-noise ratios as $\text{SNR}_j = \frac{\bar{P}_j}{N_j B}$.

## B. Effective Capacity

We first state the following result from [3], which identifies the QoS exponent for given arrival and departure processes under certain conditions.

*Theorem 1 ([3]):* Consider a queueing system, and suppose that the queue is stable and that both the arrival process $a[n], n = 1, 2, \ldots$ and service process $c[n], n = 1, 2, \ldots$ satisfy the Gärtner-Ellis limit, i.e., for all $\theta \geq 0$, there exists a differentiable asymptotic logarithmic moment generating function (LMGF) $\Lambda_A(\theta)$ defined as[1]

$$\Lambda_A(\theta) = \lim_{n \to \infty} \frac{\log \mathbb{E}\{e^{\theta \sum_{i=1}^n a[i]}\}}{n}, \qquad (6)$$

---

[1]Throughout the text, logarithm expressed without a base, i.e., $\log(\cdot)$, refers to the natural logarithm $\log_e(\cdot)$.

and a differentiable asymptotic LMGF $\Lambda_C(\theta)$ defined as

$$\Lambda_C(\theta) = \lim_{n \to \infty} \frac{\log \mathbb{E}\{e^{\theta \sum_{i=1}^n c[i]}\}}{n}. \tag{7}$$

If there exists a unique $\theta^* > 0$ such that

$$\Lambda_A(\theta^*) + \Lambda_C(-\theta^*) = 0, \tag{8}$$

then

$$\lim_{Q_{\max} \to \infty} \frac{\log \Pr\{Q > Q_{\max}\}}{Q_{\max}} = -\theta^*. \tag{9}$$

where $Q$ is the stationary queue length. ■

Now, we discuss the implications of this result on the two-hop link we study. Assume that the constant arrival rate at the source is $R \geq 0$, and the channels operate at their capacities. To satisfy the QoS constraint at the source, we should have

$$\tilde{\theta} \geq \theta_1 \tag{10}$$

where $\tilde{\theta}$ is the solution to

$$R = -\frac{\Lambda_{sr}(-\tilde{\theta})}{\tilde{\theta}} \tag{11}$$

and $\Lambda_{sr}(\theta)$ is the LMGF of the instantaneous capacity of the $\mathbf{S} - \mathbf{R}$ link.

According to [3], the LMGF of the departure process from the source, or equivalently the arrival process to the relay node, is given by

$$\Lambda_r(\theta) = \begin{cases} R\theta, & 0 \leq \theta \leq \tilde{\theta} \\ R\tilde{\theta} + \Lambda_{sr}(\theta - \tilde{\theta}), & \theta > \tilde{\theta} \end{cases}. \tag{12}$$

Therefore, in order to satisfy the QoS of the intermediate relay node $\mathbf{R}$, we must have

$$\hat{\theta} \geq \theta_2 \tag{13}$$

where $\hat{\theta}$ is the solution to

$$\Lambda_r(\hat{\theta}) + \Lambda_{rd}(-\hat{\theta}) = 0. \tag{14}$$

Above, $\Lambda_{rd}(\theta)$ is the LMGF of the instantaneous capacity of the $\mathbf{R} - \mathbf{D}$ link.

After these characterizations, effective capacity of the two-hop communication model can be formulated as follows.

*Definition 1:* The effective capacity of the two-hop communication link with the QoS constraints specified by $\theta_1$ at the source and $\theta_2$ at the relay node is given by

$$R_E(\theta_1, \theta_2) = \sup_{R \in \mathcal{R}} R \tag{15}$$

where $\mathcal{R}$ is the collection of constant arrival rates $R$ for which the solutions $\tilde{\theta}$ and $\hat{\theta}$ of (11) and (14) satisfy $\tilde{\theta} \geq \theta_1$ and $\hat{\theta} \geq \theta_2$, respectively. Hence, effective capacity is the maximum constant arrival rate that can be supported by the two-hop link in the presence of QoS constraints at both the source and relay nodes.

III. EFFECTIVE CAPACITY OF A TWO-HOP LINK IN BLOCK FADING CHANNELS

We assume that the channel state information of the links $\mathbf{S} - \mathbf{R}$ and $\mathbf{R} - \mathbf{D}$ is available at $\mathbf{S}$ and $\mathbf{R}$, and the channel state information of the link $\mathbf{R} - \mathbf{D}$ is available at $\mathbf{R}$ and $\mathbf{D}$. The transmission power levels at the source and the intermediate-hop node are fixed and hence no power control is employed (i.e., nodes are subject to short-term power constraints). We further assume that the channel capacity for each link can be achieved, i.e., the service processes are equal to the instantaneous Shannon capacities of the links. Moreover, we consider a block fading scenario in which the fading stays constant for a block of $T$ seconds and change independently from one block to another.

## A. Full-Duplex Relay

In this part, we consider the full-duplex relay. The instantaneous capacities of the $\mathbf{S}-\mathbf{R}$ and $\mathbf{R}-\mathbf{D}$ links in each block are given, respectively, by

$$TB\log_2(1+\text{SNR}_1 z_1) \quad \text{and} \quad TB\log_2(1+\text{SNR}_2 z_2) \tag{16}$$

in the units of bits per block or equivalently bits per $T$ seconds. These can be regarded as the service processes at the source and relay.

Under the block fading assumption, the logarithmic moment generating functions for the service processes of links $\mathbf{S}-\mathbf{R}$ and $\mathbf{R}-\mathbf{D}$ as functions of $\theta$ are given by[2] [6]

$$\Lambda_{sr}(\theta) = \log \mathbb{E}_{z_1}\left\{e^{\theta TB\log_2(1+\text{SNR}_1 z_1)}\right\} \tag{17}$$

$$\Lambda_{rd}(\theta) = \log \mathbb{E}_{z_2}\left\{e^{\theta TB\log_2(1+\text{SNR}_2 z_2)}\right\} \tag{18}$$

and as a result

$$\Lambda_r(\theta) = \begin{cases} R\theta, & 0 \leq \theta \leq \tilde{\theta} \\ R\tilde{\theta} + \log \mathbb{E}_{z_1}\left\{e^{(\theta-\tilde{\theta})TB\log_2(1+\text{SNR}_1 z_1)}\right\}, & \theta > \tilde{\theta} \end{cases}. \tag{19}$$

With these formulations for $\Lambda_{sr}$, $\Lambda_{rd}$, and $\Lambda_r$, we can now more explicitly express the equations in (11) and (14) as

$$R = g(\tilde{\theta}) = -\frac{1}{\tilde{\theta}}\log \mathbb{E}_{z_1}\left\{e^{-\tilde{\theta}TB\log_2(1+\text{SNR}_1 z_1)}\right\} \tag{20}$$

and

$$R = h(\tilde{\theta}, \hat{\theta}) = \begin{cases} -\frac{1}{\hat{\theta}}\log \mathbb{E}_{z_2}\left\{e^{-\hat{\theta}TB\log_2(1+\text{SNR}_2 z_2)}\right\} & 0 \leq \hat{\theta} \leq \tilde{\theta} \\ -\frac{1}{\hat{\theta}}\left(\log \mathbb{E}_{z_2}\left\{e^{-\hat{\theta}TB\log_2(1+\text{SNR}_2 z_2)}\right\} + \log \mathbb{E}_{z_1}\left\{e^{(\hat{\theta}-\tilde{\theta})TB\log_2(1+\text{SNR}_1 z_1)}\right\}\right) & \hat{\theta} \geq \tilde{\theta} \end{cases}, \tag{21}$$

---

[2]Due to the assumption that the fading changes independently from one block to another, we can, for instance, simplify (6) as $\Lambda_A = \lim_{n\to\infty}\frac{\log \mathbb{E}\{e^{\theta \sum_{i=1}^n a[i]}\}}{n} = \lim_{n\to\infty}\frac{\log \prod_{i=1}^n \mathbb{E}\{e^{\theta a[i]}\}}{n} = \lim_{n\to\infty}\frac{\sum_{i=1}^n \log \mathbb{E}\{e^{\theta a[i]}\}}{n} = \lim_{n\to\infty}\frac{n\log \mathbb{E}\{e^{\theta a[1]}\}}{n} = \log \mathbb{E}\{e^{\theta a[1]}\}.$

respectively.

We seek to identify the constant arrival rates $R$ that can be supported in the presence of QoS constraints specified by the QoS exponents $\theta_1$ for the $\mathbf{S}-\mathbf{R}$ link and $\theta_2$ for the $\mathbf{R}-\mathbf{D}$ link. In this quest, we have the following characterization. The rates $R$, which simultaneously satisfy the equations in (20) and (21) with some $\tilde{\theta} \geq \theta_1$ and $\hat{\theta} \geq \theta_2$, are the arrival rates that can be supported by the two-hop link while having the buffer violation probabilities, for large $Q_{\max}$, behave approximately as $\Pr\{Q_s \geq Q_{\max}\} \approx e^{-\tilde{\theta} Q_{\max}} \leq e^{-\theta_1 Q_{\max}}$ and $\Pr\{Q_r \geq Q_{\max}\} \approx e^{-\hat{\theta} Q_{\max}} \leq e^{-\theta_2 Q_{\max}}$, where $Q_s$ and $Q_r$ are the stationary queue lengths at the source and relay, respectively. We first establish an upper bound on these arrival rates.

*Proposition 1:* The constant arrival rates, which can be supported by the two-hop link in the presence of QoS constraints with QoS exponents $\theta_1$ and $\theta_2$ at the source and relay, respectively, are upper bounded by

$$R \leq \min\left\{-\frac{1}{\theta_1}\log \mathbb{E}_{z_1}\left\{e^{-\theta_1 TB \log_2(1+\text{SNR}_1 z_1)}\right\}, -\frac{1}{\theta_2}\log \mathbb{E}_{z_2}\left\{e^{-\theta_2 TB \log_2(1+\text{SNR}_2 z_2)}\right\}\right\}. \quad (22)$$

*Proof:* We can see from (10) and (20) that

$$R = -\frac{1}{\tilde{\theta}}\log \mathbb{E}_{z_1}\left\{e^{-\tilde{\theta} TB \log_2(1+\text{SNR}_1 z_1)}\right\} \leq -\frac{1}{\theta_1}\log \mathbb{E}_{z_1}\left\{e^{-\theta_1 TB \log_2(1+\text{SNR}_1 z_1)}\right\}. \quad (23)$$

Note that the inequality above follows from the assumption that $\tilde{\theta} \geq \theta_1$ and the fact that $-\frac{\Lambda(-\tilde{\theta})}{\tilde{\theta}} = -\frac{1}{\tilde{\theta}}\log \mathbb{E}_{z_1}\left\{e^{-\tilde{\theta} TB \log_2(1+\text{SNR}_1 z_1)}\right\}$ is a decreasing function of $\tilde{\theta}$ since larger $\tilde{\theta}$ implies a faster decay in the buffer violation probabilities and hence more stringent QoS constraints. Another upper bound can be obtained through the following arguments. Consider the idealistic scenario in which the $\mathbf{S}-\mathbf{R}$ link is deterministic (i.e., there is no fading) and can support any constant arrival rate $R$ (i.e., the capacity of this link is unbounded and $\mathbf{R}-\mathbf{D}$ link is the bottleneck). In such a case, the arriving data can immediately be sent without waiting and consequently there is no need for buffering at the source. Hence, any source QoS constraint can be satisfied. More specifically, if the service rate matches the constant arrival rate, the equation in (11) holds for any $\tilde{\theta}$, i.e.,

$$R = -\frac{\Lambda_{sr}(-\tilde{\theta})}{\tilde{\theta}} = -\frac{1}{\tilde{\theta}}\log \mathbb{E}\left\{e^{-\tilde{\theta}R}\right\} = -\frac{1}{\tilde{\theta}}(-\tilde{\theta}R) = R \quad (24)$$

where instantaneous service rate is assumed to be equal to the constant arrival rate $R$ (rather than the random quantity $TB \log_2(1+\text{SNR}_1 z_1)$ as we have in the fading channel case). Since no buffering is now required at the source, we can freely impose the most strict QoS constraints and assume $\tilde{\theta}$ to be unbounded as well. Then, we have $\hat{\theta} \leq \tilde{\theta}$ for any $\hat{\theta}$. With this, we see from (21) that

$$R = -\frac{1}{\hat{\theta}} \log \mathbb{E}_{z_2} \left\{ e^{-\hat{\theta} TB \log_2(1+\text{SNR}_2 z_2)} \right\} \leq -\frac{1}{\theta_2} \log \mathbb{E}_{z_2} \left\{ e^{-\theta_2 TB \log_2(1+\text{SNR}_2 z_2)} \right\} \quad (25)$$

where, similarly as before, the inequality is due to the assumption that $\hat{\theta} \geq \theta_2$. Combining the bounds in (23) and (25), we can equivalently write

$$R \leq \min \left\{ -\frac{1}{\theta_1} \log \mathbb{E}_{z_1} \left\{ e^{-\theta_1 TB \log_2(1+\text{SNR}_1 z_1)} \right\}, -\frac{1}{\theta_2} \log \mathbb{E}_{z_2} \left\{ e^{-\theta_2 TB \log_2(1+\text{SNR}_2 z_2)} \right\} \right\} \quad (26)$$

concluding the proof. ∎

*Remark 1:* Note that $-\frac{1}{\theta_1} \log \mathbb{E}_{z_1} \left\{ e^{-\theta_1 TB \log_2(1+\text{SNR}_1 z_1)} \right\}$ is the effective capacity of the $\mathbf{S} - \mathbf{R}$ link with QoS exponent $\theta_1$. Similarly, $-\frac{1}{\theta_2} \log \mathbb{E}_{z_2} \left\{ e^{-\theta_2 TB \log_2(1+\text{SNR}_2 z_2)} \right\}$ is the effective capacity of the $\mathbf{R} - \mathbf{D}$ link with QoS exponent $\theta_2$. Hence, the arrival rates that can be supported by the two-hop link are upper bounded by the minimum of the effective capacities of the individual links.

Below, we identify, for full-duplex relaying, the effective capacity of the two-hop link, i.e., maximum of the arrival rates that can be supported in the two-hop link in the presence of QoS constraints. According to [3], we know that the queues are not stable if the average transmission rate of link $\mathbf{R} - \mathbf{D}$ is less than the average transmission rate of link $\mathbf{S} - \mathbf{R}$. Therefore, in order to ensure stability, we assume that the condition $\mathbb{E}_{z_1}\{\log_2(1+\text{SNR}_1 z_1)\} < \mathbb{E}_{z_2}\{\log_2(1+\text{SNR}_2 z_2)\}$ is satisfied in the following result.

*Theorem 2:* The effective capacity of the two-hop communication system as a function of $\theta_1$ and $\theta_2$ is given by the following:

**Case I**: If $\theta_1 \geq \theta_2$,

$$R_E(\theta_1, \theta_2) = \min \left\{ -\frac{1}{\theta_1} \log \mathbb{E}_{z_1} \left\{ e^{-\theta_1 TB \log_2(1+\text{SNR}_1 z_1)} \right\}, -\frac{1}{\theta_2} \log \mathbb{E}_{z_2} \left\{ e^{-\theta_2 TB \log_2(1+\text{SNR}_2 z_2)} \right\} \right\}. \quad (27)$$

**Case II**: If $\theta_1 < \theta_2$ and $\theta_2 \leq \bar{\theta}$,

$$R_E(\theta_1, \theta_2) = -\frac{1}{\theta_1} \log \mathbb{E}_{z_1} \left\{ e^{-\theta_1 TB \log_2(1+\text{SNR}_1 z_1)} \right\} \tag{28}$$

where $\bar{\theta}$ is the unique value of $\theta$ for which we have the following equality satisfied:

$$-\frac{1}{\theta_1} \log \mathbb{E}_{z_1} \left\{ e^{-\theta_1 TB \log_2(1+\text{SNR}_1 z_1)} \right\} = -\frac{1}{\theta_1} \bigg( \log \mathbb{E}_{z_2} \left\{ e^{-\theta TB \log_2(1+\text{SNR}_2 z_2)} \right\} \\ + \log \mathbb{E}_{z_1} \left\{ e^{(\theta-\theta_1) TB \log_2(1+\text{SNR}_1 z_1)} \right\} \bigg). \tag{29}$$

**Case III**: Assume $\theta_1 < \theta_2$ and $\theta_2 > \bar{\theta}$.

**III.a**: If $-\frac{1}{\theta_2} \log \mathbb{E}_{z_2} \left\{ e^{-\theta_2 TB \log_2(1+\text{SNR}_2 z_2)} \right\} \geq -\frac{1}{\theta_2} \log \mathbb{E}_{z_1} \left\{ e^{-\theta_2 TB \log_2(1+\text{SNR}_1 z_1)} \right\}$, then

$$R_E(\theta_1, \theta_2) = -\frac{1}{\tilde{\theta}^*} \log \mathbb{E}_{z_1} \left\{ e^{-\tilde{\theta}^* TB \log_2(1+\text{SNR}_1 z_1)} \right\} \tag{30}$$

where $\tilde{\theta}^*$ is the smallest solution to

$$-\frac{1}{\tilde{\theta}} \log \mathbb{E}_{z_1} \left\{ e^{-\tilde{\theta} TB \log_2(1+\text{SNR}_1 z_1)} \right\} = -\frac{1}{\tilde{\theta}} \bigg( \log \mathbb{E}_{z_2} \left\{ e^{-\theta_2 TB \log_2(1+\text{SNR}_2 z_2)} \right\} \\ + \log \mathbb{E}_{z_1} \left\{ e^{(\theta_2-\tilde{\theta}) TB \log_2(1+\text{SNR}_1 z_1)} \right\} \bigg). \tag{31}$$

**III.b**: If $-\frac{1}{\theta_2} \log \mathbb{E}_{z_2} \left\{ e^{-\theta_2 TB \log_2(1+\text{SNR}_2 z_2)} \right\} < -\frac{1}{\theta_2} \log \mathbb{E}_{z_1} \left\{ e^{-\theta_2 TB \log_2(1+\text{SNR}_1 z_1)} \right\}$ and $-\frac{1}{\theta_2} \log \mathbb{E}_{z_2} \left\{ e^{-\theta_2 TB \log_2(1+\text{SNR}_2 z_2)} \right\} \geq TB \log_2(1 + \text{SNR}_1 z_{1,\min})$,

$$R_E(\theta_1, \theta_2) = -\frac{1}{\tilde{\theta}^*} \log \mathbb{E}_{z_1} \left\{ e^{-\tilde{\theta}^* TB \log_2(1+\text{SNR}_1 z_1)} \right\} \tag{32}$$

where $z_{1,\min}$ is the essential infimum of $z_1$, and $\tilde{\theta}^*$ is the solution to

$$-\frac{1}{\tilde{\theta}} \log \mathbb{E}_{z_1} \left\{ e^{-\tilde{\theta} TB \log_2(1+\text{SNR}_1 z_1)} \right\} = -\frac{1}{\theta_2} \log \mathbb{E}_{z_2} \left\{ e^{-\theta_2 TB \log_2(1+\text{SNR}_2 z_2)} \right\}. \tag{33}$$

**III.c**: Otherwise,

$$R_E(\theta_1, \theta_2) = -\frac{1}{\theta_2} \log \mathbb{E}_{z_2} \left\{ e^{-\theta_2 TB \log_2(1+\text{SNR}_2 z_2)} \right\}. \tag{34}$$

*Proof*: See Appendix A.

*Remark 2:* We see that in Case I in which $\theta_1 \geq \theta_2$, the effective capacity upper bound identified in Proposition 1 is attained.

*Remark 3:* Note that if $\theta_1 \geq \theta_2$, then the source is operating under more stringent QoS constraints then the relay. In this case, if we have

$$-\frac{1}{\theta_1} \log \mathbb{E}_{z_1} \left\{ e^{-\theta_1 TB \log_2(1+\text{SNR}_1 z_1)} \right\} \leq -\frac{1}{\theta_2} \log \mathbb{E}_{z_2} \left\{ e^{-\theta_2 TB \log_2(1+\text{SNR}_2 z_2)} \right\}, \tag{35}$$

then

$$R_E(\theta_1, \theta_2) = -\frac{1}{\theta_1} \log \mathbb{E}_{z_1} \left\{ e^{-\theta_1 TB \log_2(1+\text{SNR}_1 z_1)} \right\}. \tag{36}$$

Therefore, under these assumptions, the effective capacity is equal to the effective capacity of the $\mathbf{S} - \mathbf{R}$ link, and the performance is not affected by the presence of the buffer constraints at the relay node $\mathbf{R}$. This is because of the fact that the effective bandwidth of the departure process from the source can be completely supported by the $\mathbf{R} - \mathbf{D}$ link when the QoS exponent imposed at the relay node $\mathbf{R}$ is smaller.

The inequality in (35) is, for instance, satisfied when $z_1$ and $z_2$ (which are the fading powers in the $\mathbf{S} - \mathbf{R}$ and $\mathbf{R} - \mathbf{D}$ links) have the same distribution, and we have $\text{SNR}_1 \leq \text{SNR}_2$. We can easily see that

$$-\frac{1}{\theta_2} \log \mathbb{E}_{z_2} \left\{ e^{-\theta_2 TB \log_2(1+\text{SNR}_2 z_2)} \right\} \geq -\frac{1}{\theta_1} \log \mathbb{E}_{z_2} \left\{ e^{-\theta_1 TB \log_2(1+\text{SNR}_2 z_2)} \right\} \tag{37}$$

$$\geq -\frac{1}{\theta_1} \log \mathbb{E}_{z_1} \left\{ e^{-\theta_1 TB \log_2(1+\text{SNR}_1 z_1)} \right\} \tag{38}$$

where (37) and (38) follow from the facts that $-\frac{1}{\theta} \log \mathbb{E}_z \left\{ e^{-\theta TB \log_2(1+\text{SNR} z)} \right\}$ is a decreasing function in $\theta$, and a increasing function in SNR. This discussion also suggests that even if the source operates under more strict buffer constraints, if the fading in the $\mathbf{R} - \mathbf{D}$ link is worse than that in the $\mathbf{S} - \mathbf{R}$ link and/or the signal-to-noise ratio of the relay is smaller, i.e., $\text{SNR}_1 \geq \text{SNR}_2$, then we can have

$$R_E(\theta_1, \theta_2) = \min\left\{-\frac{1}{\theta_1}\log \mathbb{E}_{z_1}\left\{e^{-\theta_1 TB \log_2(1+\text{SNR}_1 z_1)}\right\}, -\frac{1}{\theta_2}\log \mathbb{E}_{z_2}\left\{e^{-\theta_2 TB \log_2(1+\text{SNR}_2 z_2)}\right\}\right\} \quad (39)$$

$$= -\frac{1}{\theta_2}\log \mathbb{E}_{z_2}\left\{e^{-\theta_2 TB \log_2(1+\text{SNR}_2 z_2)}\right\}, \quad (40)$$

and hence experience the $\mathbf{R} - \mathbf{D}$ link as the bottleneck.

*B. Half-Duplex Relay*

In the case of half-duplex relaying with a fixed time-sharing parameter $\tau \in (0,1)$, we assume that the source first transmits in the $\tau$ fraction of the block of $T$ seconds during which the relay listens. Subsequently, in the remaining $(1-\tau)$ fraction of the time, the relay transmits to the destination. Hence, the transmission or service rates (in bits per $T$ seconds) at the source and relay become

$$\tau TB \log_2(1 + \text{SNR}_1 z_1) \quad \text{and} \quad (1-\tau)TB \log_2(1 + \text{SNR}_2 z_2). \quad (41)$$

Now, the logarithmic moment generating functions for the service processes of links $\mathbf{S} - \mathbf{R}$ and $\mathbf{R} - \mathbf{D}$ as functions of $\theta$ are given by

$$\Lambda_{sr}(\theta) = \log \mathbb{E}_{z_1}\left\{e^{\tau\theta TB \log_2(1+\text{SNR}_1 z_1)}\right\} \quad (42)$$

$$\Lambda_{rd}(\theta) = \log \mathbb{E}_{z_2}\left\{e^{(1-\tau)\theta TB \log_2(1+\text{SNR}_2 z_2)}\right\} \quad (43)$$

and as a result, we have

$$\Lambda_r(\theta) = \begin{cases} R\theta, & 0 \leq \theta \leq \tilde{\theta} \\ R\tilde{\theta} + \log \mathbb{E}_{z_1}\left\{e^{\tau(\theta-\tilde{\theta})TB \log_2(1+\text{SNR}_1 z_1)}\right\}, & \theta > \tilde{\theta} \end{cases}.$$

With these expressions, equations in (11) and (14) can be written, for fixed $\tau$, as

$$R = g(\tilde{\theta}) = -\frac{1}{\tilde{\theta}}\log \mathbb{E}_{z_1}\left\{e^{-\tau\tilde{\theta} TB \log_2(1+\text{SNR}_1 z_1)}\right\} \quad (44)$$

and

$$R = h(\tilde{\theta}, \hat{\theta}) = \begin{cases} -\frac{1}{\hat{\theta}} \log \mathbb{E}_{z_2}\left\{e^{-(1-\tau)\hat{\theta}TB\log_2(1+\text{SNR}_2 z_2)}\right\} & 0 \leq \hat{\theta} \leq \tilde{\theta} \\ -\frac{1}{\tilde{\theta}}\left(\log \mathbb{E}_{z_2}\left\{e^{-(1-\tau)\hat{\theta}TB\log_2(1+\text{SNR}_2 z_2)}\right\} + \log \mathbb{E}_{z_1}\left\{e^{\tau(\hat{\theta}-\tilde{\theta})TB\log_2(1+\text{SNR}_1 z_1)}\right\}\right) & \hat{\theta} \geq \tilde{\theta} \end{cases},$$

(45)

respectively. As in full-duplex relaying, the rates $R$ for which the equations in (44) and (45) are simultaneously satisfied for some $\tilde{\theta} \geq \theta_1$ and $\hat{\theta} \geq \theta_2$ are the rates that can be supported by the two-hop link in the presence of QoS constraints specified by $\theta_1$ and $\theta_2$. The following result provides the effective capacity, which is defined as the supremum of such rates. Similarly as in full-duplex relaying, we assume that the average transmission rate of the $\mathbf{S} - \mathbf{R}$ link is less than the average transmission rate of the $\mathbf{R} - \mathbf{D}$ link in order to ensure stability in the buffers. Therefore, we suppose $\mathbb{E}_{z_1}\{\tau \log_2(1+\text{SNR}_1 z_1)\} < \mathbb{E}_{z_2}\{(1-\tau)\log_2(1+\text{SNR}_2 z_2)\}$. Accordingly, in the following result, we assume that the feasible values of $\tau$ for half-duplex relaying are upper bounded by

$$\tau < \tau_0 = \frac{\mathbb{E}_{z_2}\{\log_2(1+\text{SNR}_2 z_2)\}}{\mathbb{E}_{z_1}\{\log_2(1+\text{SNR}_1 z_1)\} + \mathbb{E}_{z_2}\{\log_2(1+\text{SNR}_2 z_2)\}}. \tag{46}$$

*Theorem 3:* In half-duplex relaying, the effective capacity of the two-hop communication link with statistical QoS constraints at the source and the intermediate relay nodes is given by

$$\textbf{Case I } \theta_1 \geq \theta_2: \quad R_E(\theta_1, \theta_2) = -\frac{1}{\theta_1} \log \mathbb{E}_{z_1}\left\{e^{-\tilde{\tau}\theta_1 TB\log_2(1+\text{SNR}_1 z_1)}\right\} \tag{47}$$

$$\textbf{Case II } \theta_1 < \theta_2: \quad R_E(\theta_1, \theta_2) = -\frac{1}{\theta_1} \log \mathbb{E}_{z_1}\left\{e^{-\hat{\tau}\theta_1 TB\log_2(1+\text{SNR}_1 z_1)}\right\} \tag{48}$$

where $\tilde{\tau} = \min\{\tau_0, \tau^*\}$ and $\tau^*$ is the solution to

$$-\frac{1}{\theta_1}\log \mathbb{E}_{z_1}\left\{e^{-\tau\theta_1 TB\log_2(1+\text{SNR}_1 z_1)}\right\} = -\frac{1}{\theta_2}\log \mathbb{E}_{z_2}\left\{e^{-(1-\tau)\theta_2 TB\log_2(1+\text{SNR}_2 z_2)}\right\} \tag{49}$$

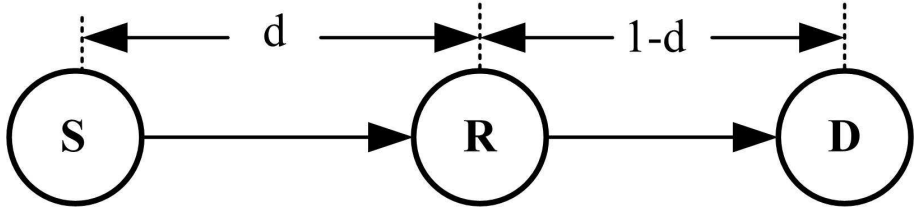

Fig. 2. The relay model.

and $\hat{\tau} = \min\{\tau_0, \tau'\}$ and $\tau'$ is the solution to

$$-\frac{1}{\theta_1} \log \mathbb{E}_{z_1} \left\{ e^{-\tau\theta_1 TB \log_2(1+\text{SNR}_1 z_1)} \right\}$$
$$= -\frac{1}{\theta_1} \left( \log \mathbb{E}_{z_2} \left\{ e^{-(1-\tau)\theta_2 TB \log_2(1+\text{SNR}_2 z_2)} \right\} + \log \mathbb{E}_{z_1} \left\{ e^{\tau(\theta_2-\theta_1)TB \log_2(1+\text{SNR}_1 z_1)} \right\} \right). \quad (50)$$

*Proof*: See Appendix B.

### C. Numerical Results

We consider the relay model depicted in Fig. 2. The source, relay, and destination nodes are located on a straight line. The distance between the source and the destination is normalized to 1. Let the distance between the source and the relay node be $d \in (0, 1)$. Then, the distance between the relay and the destination is $1-d$. We assume the fading distributions for $\mathbf{S}-\mathbf{R}$ and $\mathbf{R}-\mathbf{D}$ links follow independent Rayleigh fading with means $\mathbb{E}\{z_1\} = 1/d^\alpha$ and $\mathbb{E}\{z_2\} = 1/(1-d)^\alpha$, respectively, where we assume that the path loss $\alpha = 4$. We assume that $\text{SNR}_1 = 0$ dB and $\theta_1 = 0.01$ in the following numerical results.

In Fig. 3, we plot the effective capacity as a function of the QoS constraints of the full-duplex relay node for different $\text{SNR}_2$ values. We fix $d = 0.5$, in which case the $\mathbf{S}-\mathbf{R}$ and $\mathbf{R}-\mathbf{D}$ links have the same channel conditions. From the figure, we can see that the effective capacity does not decrease for a certain range of $\theta_2$, and this range is increased by increasing $\text{SNR}_2$. Motivated by this observation, we plot the value of $\theta_2'$, up to which the effective capacity is unaffected, as a function of $\text{SNR}_2$ in Fig. 4. Note that for all values of the pair $(\text{SNR}, \theta_2)$ below the curve shown in the figure, the QoS constraints of the relay node do not impose any negative effect on the effective capacity. This provides us with useful insight on the design of wireless systems. In Fig. 5, we plot the effective capacity as $d$ varies. We assume $\theta_2 = \{0.001, 0.01, 0.05, 0.1\}$. We are interested in the range in which the condition for stable queues (as stated above Theorem 2) is satisfied. More specifically, we note that the optimal $d$ is lower bounded by the value at which we have $\mathbb{E}_{z_1}\{\log_2(1 + \text{SNR}_1 z_1)\} = \mathbb{E}_{z_2}\{\log_2(1 + \text{SNR}_2 z_2)\}$. We can see from the figure that for small $\theta_2$ (i.e., for $\theta_2 = 0.001$ and $\theta_2 = 0.01$), the effective capacity curves overlap. In these cases, $\mathbf{S} - \mathbf{R}$ link is the bottleneck and the throughput is determined by the effective capacity of this link. When $\theta_2$ is greater than $\theta_1$ (i.e., when $\theta_2 = 0.05$ or $0.1$), it is interesting that the effective capacity decreases first and then increases until the $\mathbf{S} - \mathbf{R}$ link becomes again the bottleneck, in which case the curves overlap. This tells us that with stringent QoS constraints at the relay, having symmetric channel conditions for the links $\mathbf{S} - \mathbf{R}$ and $\mathbf{R} - \mathbf{D}$, i.e., having $d = 0.5$, generally leads to lower performance.

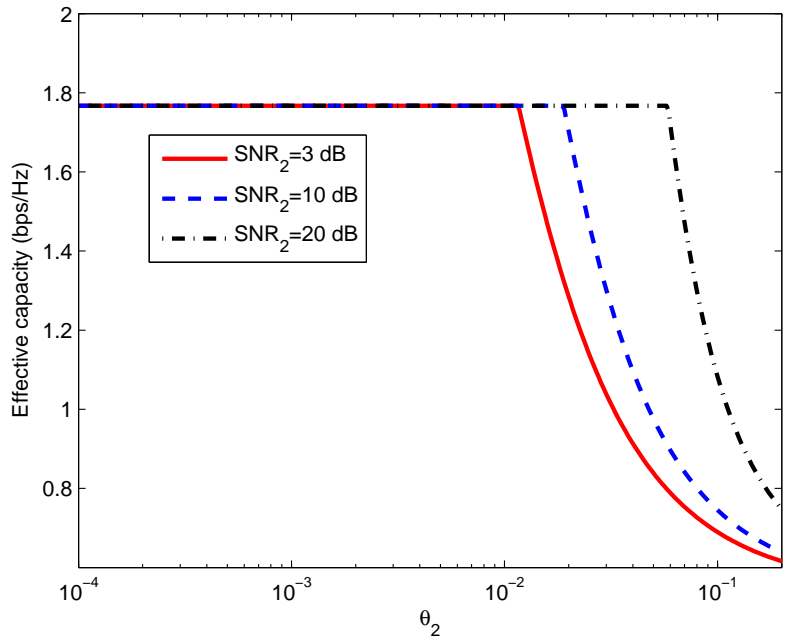

Fig. 3. The effective capacity as a function of $\theta_2$. $d = 0.5$.

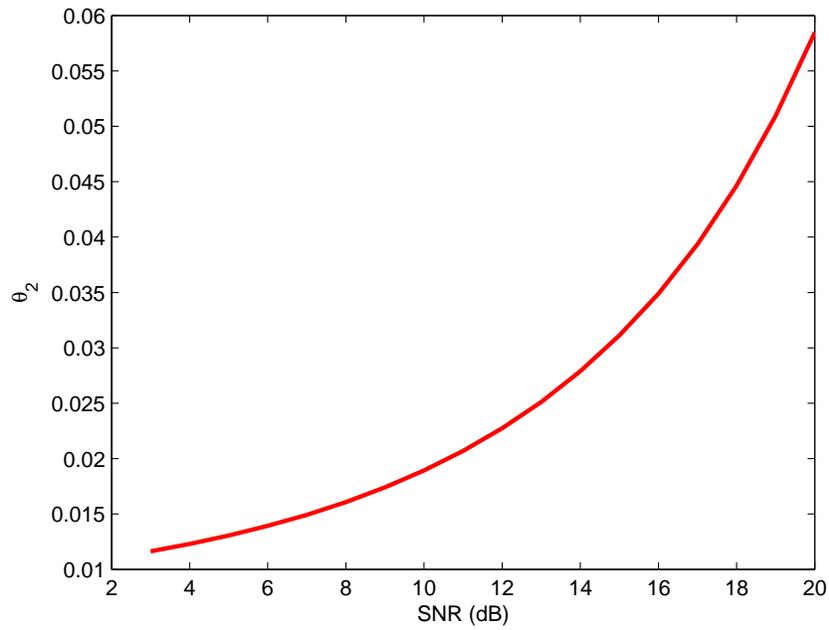

Fig. 4. $\theta'_2$ vs. $\text{SNR}_2$ for $d = 0.5$.

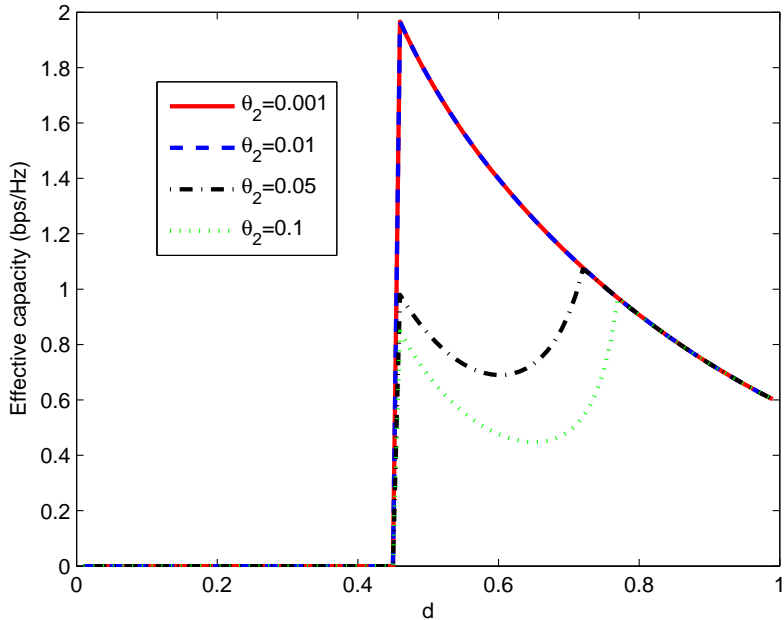

Fig. 5. The effective capacity as a function of $d$.

In Fig. 6, we plot the effective capacity as a function of $\theta_2$ for half-duplex relaying. We set $d = 0.5$. From the figure, we can find that the effective capacity stays constant for small $\theta_2$, i.e., the QoS constraints at the relay node does not impose any negative effect on the effective capacity of the system. We can also see that as $\text{SNR}_2$ increases, larger QoS constraints at the relay node can be supported while having the effective capacity of the system unaltered. One stark difference from the full-duplex relay is that as $\text{SNR}_2$ increases, the effective capacity of the system increases as well even for small $\theta_2$. This is due to the nature of the half-duplex operation. As $\text{SNR}_2$ increases, more time can be allocated to the transmission between the source and relay nodes while satisfying (46).

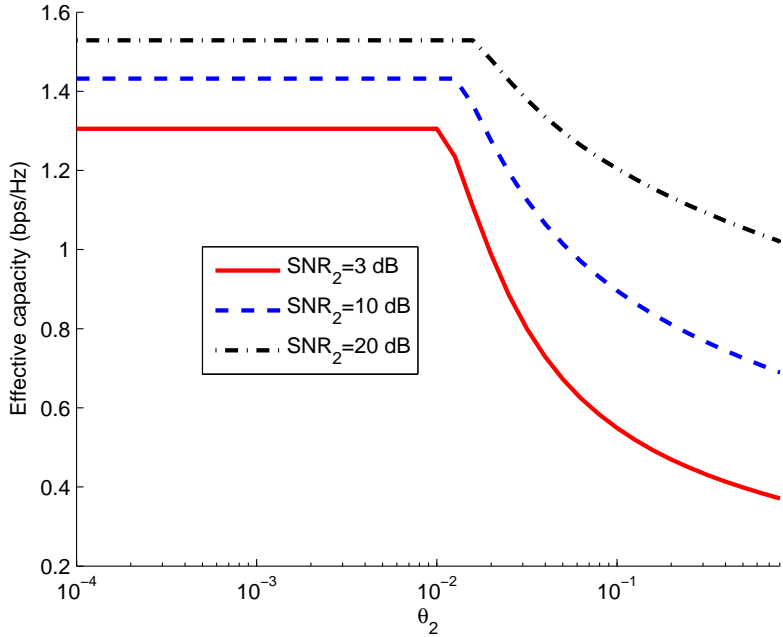

Fig. 6. The effective capacity as a function of $\theta_2$. $d = 0.5$. $\text{SNR}_2 = \{3, 10, 20\}$ dB.

In Fig. 7, we plot the effective capacity as $d$ and $\theta_2$ varies. We assume $\text{SNR}_2 = 3$ dB. As we can see from the figure, there exists an optimal $d$ that maximizes the effective capacity of the system. Besides, the optimal $d$ increases as $\theta_2$ increases. This is due to the fact that as the QoS constraints at the relay node become more stringent, the effective bandwidth supported by the $\mathbf{R} - \mathbf{D}$ link decreases and this link becomes the bottleneck of the system. In order to counterbalance this negative effect, the channel conditions of the $\mathbf{R} - \mathbf{D}$ link should be improved, which results in a larger $d$. It is also interesting that the curve is nearly flat for small $\theta_2$ when $d$ is large. So, we plot the effective capacity as $d$ varies for $\theta_2 = \{0.001, 0.01, 0.1\}$ in Fig. 8. Confirming the observation in Fig. 7, we see that the two curves for $\theta_2 = 0.001$ and $\theta_2 = 0.01$ overlap as $d$ increases. This is because the upperbound for $\tau$ specified in (46) is achieved for both curves.

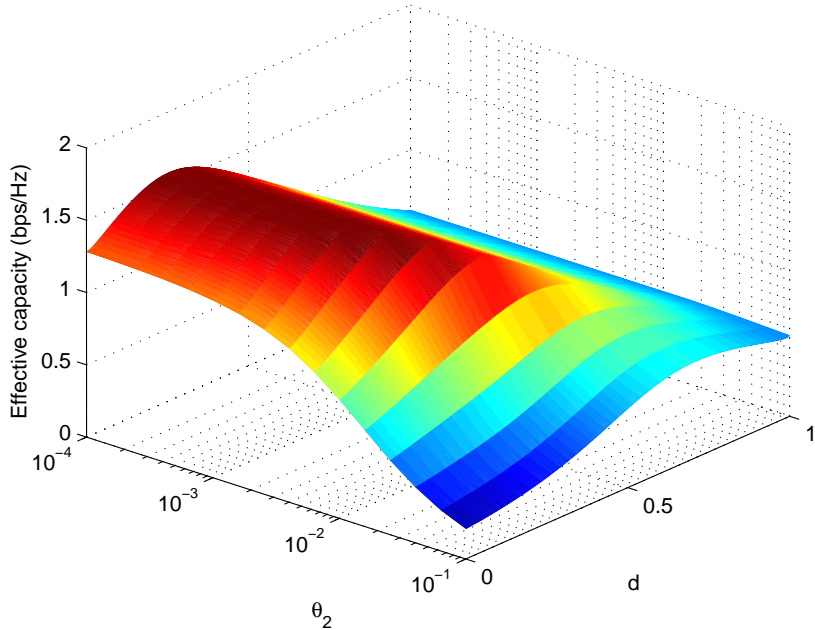

Fig. 7. The effective capacity v.s. $d$ and $\theta_2$. $\text{SNR}_2 = 3$ dB.

## IV. CONCLUSION

In this paper, we have analyzed the maximum arrival rates that can be supported by a two-hop communication link in which the source and relay nodes are both subject to statistical QoS constraints. We have determined the effective capacity in the block-fading scenario as a function of the signal-to-noise ratio levels $\text{SNR}_1$ and $\text{SNR}_2$ and the QoS exponents $\theta_1$ and $\theta_2$ for both full-duplex and half-duplex relaying. Through this analysis, we have quantified the throughput of a two-hop link operating under buffer constraints. In particular, we have shown that effective capacity can have different characterizations depending on how buffer constraints at the source and relay or more specifically how $\theta_1$ and $\theta_2$ compare. We have noted that if $\theta_1 \geq \theta_2$, the upper bound on the effective capacity is attained. We have also seen that under certain conditions depending on the SNR levels and fading distributions, $\mathbf{S} - \mathbf{R}$ link becomes the bottleneck and buffer constraints at the relay do not incur performance losses when the QoS exponent $\theta_2$ is sufficiently small but nonzero. In the numerical results, the threshold for $\theta_2$ above which the effective capacity starts diminishing is identified and is shown to increase with increasing $\text{SNR}_2$. In a simple linear setting, we have numerically investigated the impact of the location of the relay on the effective capacity for different values of the QoS exponents. In half-duplex relaying, we have determined the optimal time-sharing parameter $\tau$. In the numerical results, we have had several interesting observations. We have shown that as the SNR level at the relay node increases, the effective capacity of the system increases for all $\theta_2$. Additionally, as the QoS constraints at the relay node become more stringent, we have observed that the effective capacity of the system can be increased by improving the channel conditions in the $\mathbf{R} - \mathbf{D}$ link through having the relay node approach the destination.

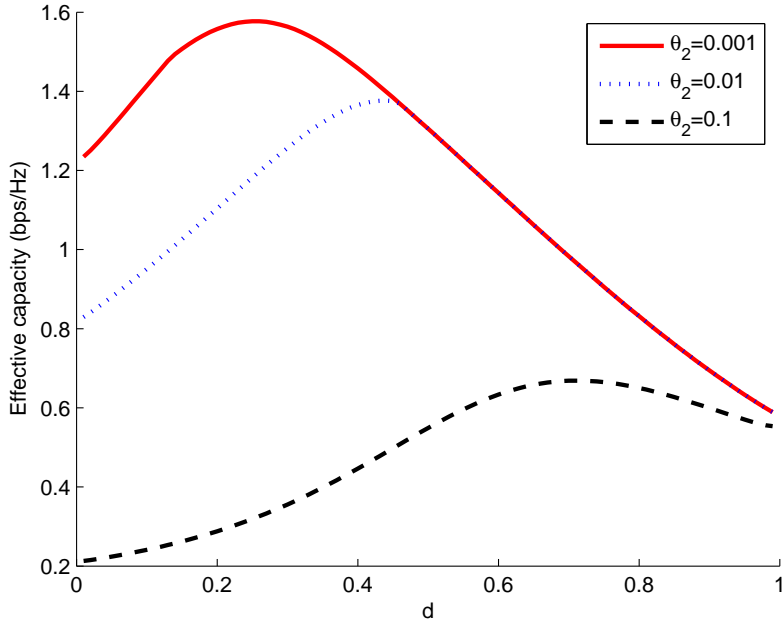

Fig. 8. The effective capacity as $d$ varies. $\text{SNR}_2 = 3$ dB. $\theta_2 = \{0.001, 0.01, 0.1\}$.

APPENDIX

## A. Proof of Theorem 2

**Case I** $\theta_1 \geq \theta_2$:

For this case, we can show that the upper bound in (22) can be attained. First assume that

$$-\frac{1}{\theta_2} \log \mathbb{E}_{z_2} \left\{ e^{-\theta_2 TB \log_2(1+\text{SNR}_2 z_2)} \right\} \leq -\frac{1}{\theta_1} \log \mathbb{E}_{z_1} \left\{ e^{-\theta_1 TB \log_2(1+\text{SNR}_1 z_1)} \right\}. \tag{51}$$

Hence, the second term on the right-hand side of (22) is the minimum one. Now, set $\hat{\theta} = \theta_2$ in (21). Assume that $\tilde{\theta} \geq \hat{\theta} = \theta_2$ where $\tilde{\theta}$ is the solution to (20). The validity of this assumption will be shown later below. Under these assumptions, we see from (21) that

$$R = h(\tilde{\theta}, \theta_2) = -\frac{1}{\theta_2} \log \mathbb{E}_{z_2} \left\{ e^{-\theta_2 TB \log_2(1+\text{SNR}_2 z_2)} \right\} \quad \text{for all } \tilde{\theta} \geq \hat{\theta} = \theta_2. \tag{52}$$

Now, in order to show that this rate can be supported, we have to prove that the equation in (20) is also satisfied for this choice of $R$, i.e., we should have

$$R = -\frac{1}{\theta_2} \log \mathbb{E}_{z_2} \left\{ e^{-\theta_2 TB \log_2(1+\text{SNR}_2 z_2)} \right\} = g(\tilde{\theta}) = -\frac{1}{\tilde{\theta}} \log \mathbb{E}_{z_1} \left\{ e^{-\tilde{\theta} TB \log_2(1+\text{SNR}_1 z_1)} \right\} \tag{53}$$

for some $\tilde{\theta}$ satisfying $\tilde{\theta} \geq \theta_1$ and $\tilde{\theta} \geq \hat{\theta} = \theta_2$. From (51) and (52), we have

$$R \leq -\frac{1}{\theta_1} \log \mathbb{E}_{z_1} \left\{ e^{-\theta_1 TB \log_2(1+\text{SNR}_1 z_1)} \right\}. \tag{54}$$

Since $-\frac{1}{\theta} \log \mathbb{E}_{z_1} \left\{ e^{-\theta TB \log_2(1+\text{SNR}_1 z_1)} \right\}$ is a decreasing function of $\theta$, (54) implies that there exists a $\tilde{\theta} \geq \theta_1$ such that

$$R = -\frac{1}{\tilde{\theta}} \log \mathbb{E}_{z_1} \left\{ e^{-\tilde{\theta} TB \log_2(1+\text{SNR}_1 z_1)} \right\} \leq -\frac{1}{\theta_1} \log \mathbb{E}_{z_1} \left\{ e^{-\theta_1 TB \log_2(1+\text{SNR}_1 z_1)} \right\} \tag{55}$$

showing that (53) holds. Note that in Case I, the original assumption is that $\theta_1 \geq \theta_2$. Then, we have $\tilde{\theta} \geq \theta_1 \geq \hat{\theta} = \theta_2$. Hence, in case I, we satisfy $\tilde{\theta} \geq \hat{\theta} = \theta_2$, verifying the earlier assumption. In summary, we have shown that (20) and (21) simultaneously hold for $\tilde{\theta} \geq \theta_1$ and $\hat{\theta} = \theta_2$ when we have

$$R = \min\left\{-\frac{1}{\theta_1}\log \mathbb{E}_{z_1}\left\{e^{-\theta_1 TB\log_2(1+\text{SNR}_1 z_1)}\right\}, -\frac{1}{\theta_2}\log \mathbb{E}_{z_2}\left\{e^{-\theta_2 TB\log_2(1+\text{SNR}_2 z_2)}\right\}\right\} \quad (56)$$

$$= -\frac{1}{\theta_2}\log \mathbb{E}_{z_2}\left\{e^{-\theta_2 TB\log_2(1+\text{SNR}_2 z_2)}\right\}. \quad (57)$$

Hence, the upper bound in (22) can be achieved and this is the effective capacity.

Above, we have assumed that the second term in (22) is the minimum one. On the other hand, if we have

$$-\frac{1}{\theta_1}\log \mathbb{E}_{z_1}\left\{e^{-\theta_1 TB\log_2(1+\text{SNR}_1 z_1)}\right\} \leq -\frac{1}{\theta_2}\log \mathbb{E}_{z_2}\left\{e^{-\theta_2 TB\log_2(1+\text{SNR}_2 z_2)}\right\}, \quad (58)$$

similar arguments follow. In particular, we can choose $\tilde{\theta} = \theta_1$ in this case, and have from (20)

$$R = g(\theta_1) = -\frac{1}{\theta_1}\log \mathbb{E}_{z_1}\left\{e^{-\theta_1 TB\log_2(1+\text{SNR}_1 z_1)}\right\}. \quad (59)$$

Through a similar approach as above, we can show that (21) can be satisfied with $\hat{\theta} \geq \theta_2$ for this choice of $R$ and establish that the upper bound in (22) is again attained.

**Case II**: $\theta_1 < \theta_2$ and $\theta_2 \leq \bar{\theta}$:

Suppose that the effective capacity is decided by the $\mathbf{S}-\mathbf{R}$ link and $\tilde{\theta} = \theta_1$ returns the highest $R$. Hence, we set $\tilde{\theta} = \theta_1$ in (20) and have

$$R = -\frac{1}{\theta_1}\log \mathbb{E}_{z_1}\left\{e^{-\theta_1 TB\log_2(1+\text{SNR}_1 z_1)}\right\}. \quad (60)$$

Clearly, this rate can be supported by the $\mathbf{S}-\mathbf{R}$ link while the QoS constraint at the source is satisfied. In order to prove that this rate is viable for the two-hop link in the presence of the QoS constraint at the relay, we have to show that the equality in (21) is satisfied as well for some $\hat{\theta} \geq \theta_2$. Note that the assumption in Case II is $\tilde{\theta} = \theta_1 < \theta_2$. Then, having $\hat{\theta} \geq \theta_2$ implies that $\hat{\theta} > \tilde{\theta} = \theta_1$. Consequently, in order to satisfy (21), we should have

$$R = -\frac{1}{\theta_1}\left(\log \mathbb{E}_{z_2}\left\{e^{-\hat{\theta} TB\log_2(1+\text{SNR}_2 z_2)}\right\} + \log \mathbb{E}_{z_1}\left\{e^{(\hat{\theta}-\theta_1)TB\log_2(1+\text{SNR}_1 z_1)}\right\}\right) \quad (61)$$

where we have used the assumption that $\tilde{\theta} = \theta_1$. Our goal is to see whether (60) and (61) for some $\hat{\theta} \geq \theta_2$ can be satisfied simultaneously. In this quest, we first show several properties of the function on the right-hand side of (61).

*Lemma 1:* Consider the function

$$f(\theta) = -\frac{1}{\theta_1}\left(\log \mathbb{E}\left\{e^{-\theta TB \log_2(1+\text{SNR}_2 z_2)}\right\} + \log \mathbb{E}\left\{e^{(\theta-\theta_1)TB \log_2(1+\text{SNR}_1 z_1)}\right\}\right) \quad \text{for } \theta \geq 0. \quad (62)$$

This function has the following properties:

a) $f(\theta)$ is a continuous function of $\theta$.

b) $f(0) = -\frac{1}{\theta_1} \log \mathbb{E}\left\{e^{-\theta_1 TB \log_2(1+\text{SNR}_1 z_1)}\right\}$.

c) The first derivative of $f(\theta)$ with respect to $\theta$ at $\theta = 0$ is positive, i.e., $\dot{f}(0) > 0$. Hence, $f(\theta)$ is initially an increasing function in the vicinity of the origin as $\theta$ increases.

d) $f(\theta)$ is a concave function of $\theta$.

e) If $TB \log_2(1 + \text{SNR}_1 z_{1,\max}) > TB \log_2(1 + \text{SNR}_2 z_{2,\min})$ where $z_{1,\max}$ is the essential supremum of the random variable $z_1$ and $z_{2,\min}$ is the essential infimum of $z_2$, then there exists a $\theta^* > 0$ such that $f(\theta^*) = 0$.

*Proof*:

a) The continuity can be shown by noting the continuity of the logarithm and exponential functions and employing the Dominated Convergence Theorem and Monotone Convergence Theorem for the justification of the interchange of the limit and expectations. For the first expectation in (62), we can apply the Dominated Convergence Theorem by observing that we have $|e^{-\theta TB \log_2(1+\text{SNR}_2 z_2)}| \leq 1$ for all $\theta \geq 0$ and the bounding function is integrable, i.e., $\mathbb{E}\{1\} = 1 < \infty$. For the second expectation, we immediately note that $e^{(\theta-\theta_1)TB \log_2(1+\text{SNR}_1 z_1)}$ is nonnegative and increases with increasing $\theta$, and consequently we can use the Monotone Convergence Theorem to justify the interchange of limit and expectation.

b) This property can be readily seen by evaluating the function at $\theta = 0$.

c) The first derivative of $f$ with respect to $\theta$ can be evaluated as

$$\dot{f}(\theta) = -\frac{1}{\theta_1}\left(\frac{-\mathbb{E}_{z_2}\left\{e^{-\theta TB\log_2(1+\text{SNR}_2 z_2)}TB\log_2(1+\text{SNR}_2 z_2)\right\}}{\mathbb{E}_{z_2}\left\{e^{-\theta TB\log_2(1+\text{SNR}_2 z_2)}\right\}} \right.$$
$$\left. + \frac{\mathbb{E}_{z_1}\left\{e^{(\theta-\theta_1)TB\log_2(1+\text{SNR}_1 z_1)}TB\log_2(1+\text{SNR}_1 z_1)\right\}}{\mathbb{E}_{z_1}\left\{e^{(\theta-\theta_1)TB\log_2(1+\text{SNR}_1 z_1)}\right\}}\right). \tag{63}$$

Then, $\dot{f}(0)$ can be written as

$$\dot{f}(0) = \frac{TB}{\theta_1}\left(\mathbb{E}_{z_2}\{\log_2(1+\text{SNR}_2 z_2)\} - \frac{\mathbb{E}_{z_1}\{e^{-\theta_1 TB\log_2(1+\text{SNR}_1 z_1)}\log_2(1+\text{SNR}_1 z_1)\}}{\mathbb{E}_{z_1}\{e^{-\theta_1 TB\log_2(1+\text{SNR}_1 z_1)}\}}\right). \tag{64}$$

Let us define

$$\alpha(\theta_1) = \mathbb{E}_{z_2}\{\log_2(1+\text{SNR}_2 z_2)\} - \frac{\mathbb{E}_{z_1}\{e^{-\theta_1 TB\log_2(1+\text{SNR}_1 z_1)}\log_2(1+\text{SNR}_1 z_1)\}}{\mathbb{E}_{z_1}\{e^{-\theta_1 TB\log_2(1+\text{SNR}_1 z_1)}\}}. \tag{65}$$

We can see that $\alpha(0) = \mathbb{E}_{z_2}\{\log_2(1+\text{SNR}_2 z_2)\} - \mathbb{E}_{z_1}\{\log_2(1+\text{SNR}_1 z_1)\} > 0$ (due to our original assumption to ensure stability). The first derivative of $\alpha(\theta_1)$ with respect to $\theta_1$ is

$$\dot{\alpha}(\theta_1) = TB\frac{1}{\left(\mathbb{E}_{z_1}\{e^{-\theta_1 TB\log_2(1+\text{SNR}_1 z_1)}\}\right)^2}$$
$$\times \left(\mathbb{E}_{z_1}\{e^{-\theta_1 TB\log_2(1+\text{SNR}_1 z_1)}(\log_2(1+\text{SNR}_1 z_1))^2\}\mathbb{E}_{z_1}\{e^{-\theta_1 TB\log_2(1+\text{SNR}_1 z_1)}\} \right.$$
$$\left. - \left(\mathbb{E}_{z_1}\{e^{-\theta_1 TB\log_2(1+\text{SNR}_1 z_1)}\log_2(1+\text{SNR}_1 z_1)\}\right)^2\right) \tag{66}$$

By Cauchy-Schwarz inequality, we know that $\mathbb{E}\{X^2\}\mathbb{E}\{Y^2\} \geq (\mathbb{E}\{XY\})^2$. Then, denoting $X = \sqrt{e^{-\theta_1 TB\log_2(1+\text{SNR}_1 z_1)}(\log_2(1+\text{SNR}_1 z_1))^2}$ and $Y = \sqrt{e^{-\theta_1 TB\log_2(1+\text{SNR}_1 z_1)}}$, we easily see that $\dot{\alpha}(\theta_1) \geq 0$ for all $\theta_1$. Thus, $\alpha(\theta_1)$ is an increasing function and we have $\alpha(\theta_1) \geq \alpha(0) > 0$. Hence, $\dot{f}(0) > 0$.

d) The second derivative of $f$ with respect to $\theta$ can be expressed as

$$\ddot{f}(\theta) = -\frac{1}{\theta_1} \Bigg( \frac{1}{(\mathbb{E}_{z_2}\{e^{-\theta TB \log_2(1+\text{SNR}_2 z_2)}\})^2}$$

$$\times \Bigg( \mathbb{E}_{z_2}\left\{e^{-\theta TB \log_2(1+\text{SNR}_2 z_2)}(TB \log_2(1+\text{SNR}_2 z_2))^2\right\} \mathbb{E}_{z_2}\left\{e^{-\theta TB \log_2(1+\text{SNR}_2 z_2)}\right\}$$

$$- \left(\mathbb{E}_{z_2}\left\{e^{-\theta TB \log_2(1+\text{SNR}_2 z_2)} TB \log_2(1+\text{SNR}_2 z_2)\right\}\right)^2 \Bigg)$$

$$+ \frac{1}{(\mathbb{E}_{z_1}\{e^{(\theta-\theta_1) TB \log_2(1+\text{SNR}_1 z_1)}\})^2}$$

$$\times \Bigg( \mathbb{E}_{z_1}\left\{e^{(\theta-\theta_1) TB \log_2(1+\text{SNR}_1 z_1)}(TB \log_2(1+\text{SNR}_1 z_1))^2\right\} \mathbb{E}_{z_1}\left\{e^{(\theta-\theta_1) TB \log_2(1+\text{SNR}_1 z_1)}\right\}$$

$$- \left(\mathbb{E}_{z_1}\left\{e^{(\theta-\theta_1) TB \log_2(1+\text{SNR}_1 z_1)} TB \log_2(1+\text{SNR}_1 z_1)\right\}\right)^2 \Bigg) \Bigg) \quad (67)$$

$$\leq 0 \quad (68)$$

where Cauchy-Schwarz inequality is used again. With this characterization, we establish that $f$ is a concave function of $\theta$.

e) We first express $f(\theta)$ in the following form:

$$f(\theta) = -\frac{1}{\theta_1}\left(\log \mathbb{E}_{z_2}\left\{e^{-\theta TB \log_2(1+\text{SNR}_2 z_2)}\right\} + \log \mathbb{E}_{z_1}\left\{e^{(\theta-\theta_1) TB \log_2(1+\text{SNR}_1 z_1)}\right\}\right) \quad (69)$$

$$= \frac{\theta}{\theta_1}\left(-\frac{1}{\theta}\log \mathbb{E}_{z_2}\left\{e^{-\theta TB \log_2(1+\text{SNR}_2 z_2)}\right\} - \left(1-\frac{\theta_1}{\theta}\right)\frac{1}{\theta-\theta_1}\log \mathbb{E}_{z_1}\left\{e^{(\theta-\theta_1) TB \log_2(1+\text{SNR}_1 z_1)}\right\}\right)$$

$$= \frac{\theta}{\theta_1}\left(E_C(\theta) - E_B(\theta-\theta_1)\right) \quad (70)$$

where

$$E_C(\theta) = -\frac{1}{\theta}\log \mathbb{E}_{z_2}\left\{e^{-\theta TB \log_2(1+\text{SNR}_2 z_2)}\right\} \quad (71)$$

is the virtual effective capacity with respect to $\theta$, and

$$E_B(\theta-\theta_1) = \left(1-\frac{\theta_1}{\theta}\right)\frac{1}{\theta-\theta_1}\log \mathbb{E}_{z_1}\left\{e^{(\theta-\theta_1) TB \log_2(1+\text{SNR}_1 z_1)}\right\}$$

is the virtual effective bandwidth with respect to $\theta - \theta_1$. Similar to the discussion in [5], we know that $E_C(\theta)$ is decreasing in $\theta$. Moreover, when $\theta = 0$, we have $E_C(0) = \mathbb{E}_{z_2}\{TB\log_2(1 + \text{SNR}_2 z_2)\}$, and as $\theta \to \infty$, $E_C(\theta)$ approaches the delay limited capacity [9], i.e., $E_C(\theta) \to TB\log_2(1 + \text{SNR}_2 z_{2,\min})$ where $z_{2,\min}$ is the essential infimum of the random variable $z_2$. Furthermore, $E_B(\theta - \theta_1)$ is an increasing function of $\theta$. For $\theta < \theta_1$, $E_B(\theta - \theta_1)$ has a negative value. At $\theta = \theta_1$, we have $E_B(\theta_1 - \theta_1) = E_B(0) = 0$. As $\theta \to \infty$, $E_B(\theta - \theta_1)$ approaches the highest rate of the $\mathbf{S} - \mathbf{R}$ link, i.e., $E_B(\theta - \theta_1) \to TB\log_2(1 + \text{SNR}_1 z_{1,\max})$ where $z_{1,\max}$ is the essential supremum of the random variable $z_1$. Therefore, as long as $TB\log_2(1 + \text{SNR}_1 z_{1,\max}) > TB\log_2(1 + \text{SNR}_2 z_{2,\min})$, the decreasing curve $E_C(\theta)$ and increasing curve $E_B(\theta - \theta_1)$ will meet at some point $\theta = \theta^* > 0$ at which we have $f(\theta^*) = \frac{\theta^*}{\theta_1}(E_C(\theta^*) - E_B(\theta^* - \theta_1)) = 0$. A numerical result provides a visualization of the above discussion. In Fig. 9, we plot the virtual effective capacity and virtual effective bandwidth normalized by $TB$ as a function of $\theta$ in the Rayleigh fading channel. We assume that $T = 2$ ms, $B = 10^5$ Hz, $\theta_1 = 0.01$, $\text{SNR}_1 = 0$ dB, and $\text{SNR}_2 = 10$ dB. Note that we have $z_{1,\max} = \infty$ and $z_{2,\min} = 0$ in the Rayleigh fading model. ∎

Recall that we are seeking to establish whether (60) and (61) can simultaneously be satisfied for some $\hat{\theta} \geq \theta_2$. With the definition of the function $f(\cdot)$ whose properties are delineated in Lemma 1, the equations in (60) and (61) can be combined to write

$$f(\hat{\theta}) = -\frac{1}{\theta_1}\log\mathbb{E}_{z_1}\left\{e^{-\theta_1 TB\log_2(1+\text{SNR}_1 z_1)}\right\}. \tag{72}$$

Hence, our goal is to see whether the equation in (72) can be satisfied for some $\hat{\theta} \geq \theta_2$. In Lemma 1, we have noted that the function $f(\theta)$ is equal to the right-hand side of (72) at $\theta = 0$, and then it increases. At some point, $f(\theta)$ approaches zero. Since it is a concave function, we immediately see that $f(\theta)$ is a function that initially increases, hits a peak value, and then starts decreasing. This leads us to conclude that $f(\theta)$ becomes equal to the right-hand side of (72) once again at some unique $\theta > 0$. Let us denote this unique point as $\bar{\theta}$. Hence,

$$f(\bar{\theta}) = -\frac{1}{\theta_1}\log\mathbb{E}_{z_1}\left\{e^{-\theta_1 TB\log_2(1+\text{SNR}_1 z_1)}\right\}. \tag{73}$$

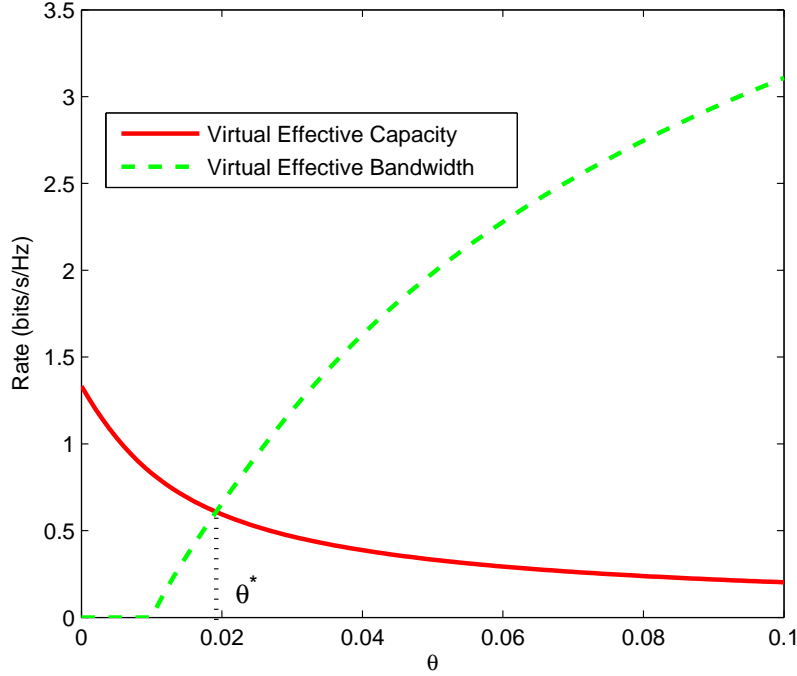

Fig. 9. The virtual effective capacity and virtual effective bandwidth as a function of $\theta$ in Rayleigh fading channels with full-duplex relay. $\mathbb{E}\{z_1\} = \mathbb{E}\{z_2\} = 1$.

If $\bar{\theta} \geq \theta_2$, then (72) is satisfied for $\hat{\theta} = \bar{\theta} \geq \theta_2$. Therefore, (60) and (61) are satisfied simultaneously. Hence, the arrival rate

$$R = -\frac{1}{\theta_1} \log \mathbb{E}_{z_1} \left\{ e^{-\theta_1 TB \log_2(1+\text{SNR}_1 z_1)} \right\} \tag{74}$$

can be supported by the two-hop link. Since this rate is an upper bound on the arrival rates as proved in Proposition 1, this arrival rate is the effective capacity, proving (28) in Theorem 2.

It is important to note that the above result implicitly assumes that $TB \log_2(1+\text{SNR}_1 z_{1,\max}) > TB \log_2(1+\text{SNR}_2 z_{2,\min})$ which is a condition in part e) of Lemma 1. Note that if this condition does not hold, then it means that the maximum service rate from the source is equal to or lower than the minimum service rate from the relay. Hence, the relay can immediately support any arrival rate without requiring any buffering. The bottleneck is the $\mathbf{S}-\mathbf{R}$ link and arrival rates are limited by the effective capacity of this link. Therefore, we again have effective capacity of the two-hop link given by (28).

**Case III**: Assume $\theta_1 < \theta_2$ and $\theta_2 > \bar{\theta}$:

Above, we have discussed the case in which $\bar{\theta} \geq \theta_2$. If, on the other hand, $\bar{\theta} < \theta_2$, then (72) and consequently (61) cannot be satisfied for some $\hat{\theta} \geq \theta_2$. Hence, the arrival rate in (74) cannot be supported by the two-hop link, and we need to consider possibly smaller rates, i.e.,

$$R = g(\tilde{\theta}) = -\frac{1}{\tilde{\theta}} \log \mathbb{E}_{z_1} \left\{ e^{-\tilde{\theta} TB \log_2(1+\text{SNR}_1 z_1)} \right\} \tag{75}$$

for some $\tilde{\theta} \geq \theta_1$. The rate given above is supported by the two-hop link if the equation

$$g(\tilde{\theta}) = h(\tilde{\theta}, \hat{\theta}) \tag{76}$$

is satisfied for some $\hat{\theta} \geq \theta_2$ and $\tilde{\theta} \geq \theta_1$. Recall that the function $h$ is defined in (21) as

$$h(\tilde{\theta}, \hat{\theta}) = \begin{cases} -\frac{1}{\tilde{\theta}} \log \mathbb{E}_{z_2} \left\{ e^{-\hat{\theta} TB \log_2(1+\text{SNR}_2 z_2)} \right\} & 0 \leq \hat{\theta} \leq \tilde{\theta} \\ -\frac{1}{\tilde{\theta}} \left( \log \mathbb{E}_{z_2} \left\{ e^{-\hat{\theta} TB \log_2(1+\text{SNR}_2 z_2)} \right\} + \log \mathbb{E}_{z_1} \left\{ e^{(\hat{\theta}-\tilde{\theta}) TB \log_2(1+\text{SNR}_1 z_1)} \right\} \right) & \hat{\theta} \geq \tilde{\theta} \end{cases}. \tag{77}$$

We first note that for fixed $\tilde{\theta}$, $h(\tilde{\theta}, \hat{\theta})$ is a decreasing function of $\hat{\theta}$ because as $\hat{\theta}$ increases, the QoS constraints at the relay become more stringent and consequently lower rates can be supported by the relay. Therefore, in order to identify the highest arrival rates $R$, we consider the smallest allowed value of $\hat{\theta}$ and set $\hat{\theta} = \theta_2$. We now consider the equation

$$g(\tilde{\theta}) = h(\tilde{\theta}, \theta_2) \tag{78}$$

and seek whether this equation is satisfied for some $\tilde{\theta} \geq \theta_1$. At $\tilde{\theta} = \theta_1$, the left-hand side of (78) becomes

$$g(\theta_1) = -\frac{1}{\theta_1} \log \mathbb{E}_{z_1} \left\{ e^{-\theta_1 TB \log_2(1+\text{SNR}_1 z_1)} \right\} \tag{79}$$

while the right-hand side is

$$h(\theta_1, \theta_2) = -\frac{1}{\theta_1} \left( \log \mathbb{E}_{z_2} \left\{ e^{-\theta_2 TB \log_2(1+\text{SNR}_2 z_2)} \right\} + \log \mathbb{E}_{z_1} \left\{ e^{(\theta_2-\theta_1) TB \log_2(1+\text{SNR}_1 z_1)} \right\} \right) \quad (80)$$

$$= f(\theta_2) \quad (81)$$

where $f(\cdot)$ is the function defined in Lemma 1. Note that our assumption in this case is $\theta_2 > \bar{\theta}$. Recalling (73), we know that

$$f(\bar{\theta}) = -\frac{1}{\theta_1} \log \mathbb{E}_{z_1} \left\{ e^{-\theta_1 TB \log_2(1+\text{SNR}_1 z_1)} \right\} = g(\theta_1). \quad (82)$$

Then, from the properties of $f$ and the assumption that $\theta_2 > \bar{\theta}$, we immediately see that

$$f(\theta_2) = h(\theta_1, \theta_2) \leq -\frac{1}{\theta_1} \log \mathbb{E}_{z_1} \left\{ e^{-\theta_1 TB \log_2(1+\text{SNR}_1 z_1)} \right\} = g(\theta_1). \quad (83)$$

Therefore, at $\tilde{\theta} = \theta_1$, the left-hand side of (78) is larger than the value at the right-hand side.

Now, let us consider the values at $\tilde{\theta} = \theta_2$. The left-hand and right-hand sides of (78) become, respectively,

$$g(\theta_2) = -\frac{1}{\theta_2} \log \mathbb{E}_{z_1} \left\{ e^{-\theta_2 TB \log_2(1+\text{SNR}_1 z_1)} \right\} \quad (84)$$

and

$$h(\theta_2, \theta_2) = -\frac{1}{\theta_2} \log \mathbb{E}_{z_2} \left\{ e^{-\theta_2 TB \log_2(1+\text{SNR}_2 z_2)} \right\} \quad (85)$$

If we have

$$g(\theta_2) = -\frac{1}{\theta_2} \log \mathbb{E}_{z_1} \left\{ e^{-\theta_2 TB \log_2(1+\text{SNR}_1 z_1)} \right\} \leq h(\theta_2, \theta_2) = -\frac{1}{\theta_2} \log \mathbb{E}_{z_2} \left\{ e^{-\theta_2 TB \log_2(1+\text{SNR}_2 z_2)} \right\}, \quad (86)$$

then the left-hand side of (78) is smaller that the value of the right-hand side at $\theta_2$. Therefore, being continuous functions, $g(\tilde{\theta})$ and $h(\tilde{\theta}, \theta_2)$ meet at some $\theta_1 \leq \tilde{\theta} \leq \theta_2$. Denote the smallest value of $\tilde{\theta}$ for which we have $g(\tilde{\theta}) = h(\tilde{\theta}, \theta_2)$ as $\tilde{\theta}^*$. Then, the highest rate that can be supported by the two-hop link is

$$R = g(\tilde{\theta}^*) = -\frac{1}{\tilde{\theta}^*} \log \mathbb{E}_{z_1}\left\{e^{-\tilde{\theta}^* TB \log_2(1+\text{SNR}_1 z_1)}\right\} \tag{87}$$

Above result is obtained under the assumption that $g(\theta_2) \leq h(\theta_2, \theta_2)$. Let us now consider the other possibility in which $g(\theta_2) > h(\theta_2, \theta_2)$. For this case, we first have the following lemma.

*Lemma 2:* Assume that $g(\theta_2) > h(\theta_2, \theta_2)$. Then, $h(\tilde{\theta}, \theta_2)$ is an increasing function of $\tilde{\theta}$ for $\tilde{\theta} \leq \theta_2$.

*Proof:* For $\tilde{\theta} \leq \theta_2$, we can express

$$h(\tilde{\theta}, \theta_2) = -\frac{1}{\tilde{\theta}} \left( \log \mathbb{E}_{z_2}\left\{e^{-\theta_2 TB \log_2(1+\text{SNR}_2 z_2)}\right\} + \log \mathbb{E}_{z_1}\left\{e^{(\theta_2-\tilde{\theta})TB \log_2(1+\text{SNR}_1 z_1)}\right\} \right). \tag{88}$$

The first derivative of $h(\tilde{\theta}, \theta_2)$ with respect to $\tilde{\theta}$ is

$$\dot{h}(\tilde{\theta}, \theta_2) = \frac{1}{\tilde{\theta}^2} \left( \tilde{\theta} \frac{\mathbb{E}_{z_1}\left\{e^{(\theta_2-\tilde{\theta})TB \log_2(1+\text{SNR}_1 z_1)} TB \log_2(1+\text{SNR}_1 z_1)\right\}}{e^{(\theta_2-\tilde{\theta})TB \log_2(1+\text{SNR}_1 z_1)}} \right.$$
$$\left. + \log \mathbb{E}_{z_1}\left\{e^{(\theta_2-\tilde{\theta})TB \log_2(1+\text{SNR}_1 z_1)}\right\} + \log \mathbb{E}_{z_2}\left\{e^{-\theta_2 TB \log_2(1+\text{SNR}_2 z_2)}\right\} \right). \tag{89}$$

Let us define

$$\beta(\tilde{\theta}) = \tilde{\theta} \frac{\mathbb{E}_{z_1}\left\{e^{(\theta_2-\tilde{\theta})TB \log_2(1+\text{SNR}_1 z_1)} TB \log_2(1+\text{SNR}_1 z_1)\right\}}{\mathbb{E}_{z_1}\left\{e^{(\theta_2-\tilde{\theta})TB \log_2(1+\text{SNR}_1 z_1)}\right\}}$$
$$+ \log \mathbb{E}_{z_1}\left\{e^{(\theta_2-\tilde{\theta})TB \log_2(1+\text{SNR}_1 z_1)}\right\} + \log \mathbb{E}_{z_2}\left\{e^{-\theta_2 TB \log_2(1+\text{SNR}_2 z_2)}\right\}. \tag{90}$$

We can show that $\beta(\tilde{\theta})$ is nonnegative.

The first derivative of $\beta(\tilde{\theta})$ with respect to $\tilde{\theta}$ is

$$\dot{\beta}(\tilde{\theta}) = \frac{\tilde{\theta}}{\left(\mathbb{E}_{z_1}\left\{e^{(\theta_2-\tilde{\theta})TB \log_2(1+\text{SNR}_1 z_1)}\right\}\right)^2} \left( -\mathbb{E}_{z_1}\left\{e^{(\theta_2-\tilde{\theta})TB \log_2(1+\text{SNR}_1 z_1)} (TB \log_2(1+\text{SNR}_1 z_1))^2\right\} \right.$$
$$\left. \times \mathbb{E}_{z_1}\left\{e^{(\theta_2-\tilde{\theta})TB \log_2(1+\text{SNR}_1 z_1)}\right\} + \left(\mathbb{E}_{z_1}\left\{e^{(\theta_2-\tilde{\theta})TB \log_2(1+\text{SNR}_1 z_1)} TB \log_2(1+\text{SNR}_1 z_1)\right\}\right)^2 \right) \tag{91}$$

$$\leq 0 \tag{92}$$

where Cauchy-Schwarz inequality is used for (92). Therefore, $\beta(\tilde{\theta})$ is a decreasing function of $\tilde{\theta}$, and hence for $\tilde{\theta} \leq \theta_2$ we have

$$\beta(\tilde{\theta}) \geq \beta(\theta_2) = \theta_2 \mathbb{E}_{z_1} \{TB \log_2(1 + \text{SNR}_1 z_1)\} + \log \mathbb{E}_{z_2} \left\{ e^{-\theta_2 TB \log_2(1+\text{SNR}_2 z_2)} \right\} \quad (93)$$

$$= -\theta_2 \left( -TB\mathbb{E}_{z_1}\{\log_2(1 + \text{SNR}_1 z_1)\} - \frac{1}{\theta_2} \log \mathbb{E}_{z_2} \left\{ e^{-\theta_2 TB \log_2(1+\text{SNR}_2 z_2)} \right\} \right) \quad (94)$$

Note that our assumption is that

$$g(\theta_2) = -\frac{1}{\theta_2} \log \mathbb{E}_{z_1} \left\{ e^{-\theta_2 TB \log_2(1+\text{SNR}_1 z_1)} \right\} > h(\theta_2, \theta_2) = -\frac{1}{\theta_2} \log \mathbb{E}_{z_2} \left\{ e^{-\theta_2 TB \log_2(1+\text{SNR}_2 z_2)} \right\}. \quad (95)$$

Since $TB\mathbb{E}_{z_1}\{\log_2(1 + \text{SNR}_1 z_1)\} \geq -\frac{1}{\theta_2} \log \mathbb{E}_{z_1} \left\{ e^{-\theta_2 TB \log_2(1+\text{SNR}_1 z_1)} \right\}$, the above inequality implies that

$$TB\mathbb{E}_{z_1}\{\log_2(1 + \text{SNR}_1 z_1)\} > -\frac{1}{\theta_2} \log \mathbb{E}_{z_2} \left\{ e^{-\theta_2 TB \log_2(1+\text{SNR}_2 z_2)} \right\} \quad (96)$$

which further implies that $\beta(\theta_2) > 0$. Finally, we immediately see that

$$\dot{h}(\tilde{\theta}, \theta_2) = \frac{1}{\tilde{\theta}^2} \beta(\tilde{\theta}) \geq \frac{1}{\tilde{\theta}^2} \beta(\theta_2) > 0 \quad (97)$$

proving that $h(\tilde{\theta}, \theta_2)$ is an increasing function of $\tilde{\theta}$ for $\tilde{\theta} \leq \theta_2$. ∎

In effect, we have shown that if $h(\theta_2, \theta_2) < g(\theta_2)$, then $h(\tilde{\theta}, \theta_2) < g(\theta_2)$ for all $\tilde{\theta} \leq \theta_2$. Note that since $g(\tilde{\theta})$ is a decreasing function, $g(\theta_2) \leq g(\tilde{\theta})$ for all $\tilde{\theta} \leq \theta_2$. Combining these, we observe that

$$h(\tilde{\theta}, \theta_2) < g(\theta_2) \leq g(\tilde{\theta}) \quad \forall \tilde{\theta} \leq \theta_2. \quad (98)$$

Therefore, the equality $g(\tilde{\theta}) = h(\tilde{\theta}, \theta_2)$ cannot be satisfied for any $\theta_1 \leq \tilde{\theta} \leq \theta_2$. Hence, we should have $\tilde{\theta} > \theta_2$. Note that for $\tilde{\theta} > \theta_2$, $h(\tilde{\theta}, \theta_2)$, which can be expressed as

$$h(\tilde{\theta}, \theta_2) = -\frac{1}{\theta_2} \log \mathbb{E}_{z_2} \left\{ e^{-\theta_2 TB \log_2(1+\text{SNR}_2 z_2)} \right\}, \quad (99)$$

is a constant for given $\theta_2$. On the other hand,

$$g(\tilde{\theta}) = -\frac{1}{\tilde{\theta}} \log \mathbb{E}_{z_1} \left\{ e^{-\tilde{\theta} TB \log_2(1+\text{SNR}_1 z_1)} \right\} \tag{100}$$

is a decreasing function with minimum value given by

$$\lim_{\tilde{\theta} \to \infty} g(\tilde{\theta}) = TB \log_2(1 + \text{SNR}_1 z_{1,\min}) \tag{101}$$

where $z_{1,\min}$ is the essential infimum of $z_1$. Hence, if

$$TB \log_2(1 + \text{SNR}_1 z_{1,\min}) \leq h(\tilde{\theta}, \theta_2) = -\frac{1}{\theta_2} \log \mathbb{E}_{z_2} \left\{ e^{-\theta_2 TB \log_2(1+\text{SNR}_2 z_2)} \right\}, \tag{102}$$

then the equation $g(\tilde{\theta}) = h(\tilde{\theta}, \theta_2)$ can be satisfied at some $\tilde{\theta} = \tilde{\theta}^* \geq \theta_2$, and the maximum arrival rate is given by

$$R = g(\tilde{\theta}^*) = -\frac{1}{\tilde{\theta}^*} \log \mathbb{E}_{z_1} \left\{ e^{-\tilde{\theta}^* TB \log_2(1+\text{SNR}_1 z_1)} \right\}. \tag{103}$$

If, on the other hand,

$$TB \log_2(1 + \text{SNR}_1 z_{1,\min}) > h(\tilde{\theta}, \theta_2) = -\frac{1}{\theta_2} \log \mathbb{E}_{z_2} \left\{ e^{-\theta_2 TB \log_2(1+\text{SNR}_2 z_2)} \right\}, \tag{104}$$

the bottleneck is the $\mathbf{R} - \mathbf{D}$ link, and the highest arrival rate that can be supported by the two-hop link is

$$R = -\frac{1}{\theta_2} \log \mathbb{E}_{z_2} \left\{ e^{-\theta_2 TB \log_2(1+\text{SNR}_2 z_2)} \right\}. \tag{105}$$

Note that this arrival rate is smaller than the smallest possible transmission rate of the source and hence no buffering is needed at the source in this extreme case. ∎

## B. Proof of Theorem 3

We first identify the following upper bound on the rates that can be supported with half-duplex relaying in the two-hop link:

$$R \leq \sup_{\tau \in [0,\tau_0)} \min\left\{-\frac{1}{\theta_1}\log \mathbb{E}_{z_1}\left\{e^{-\tau\theta_1 TB\log_2(1+\text{SNR}_1 z_1)}\right\}, -\frac{1}{\theta_2}\log \mathbb{E}_{z_2}\left\{e^{-(1-\tau)\theta_2 TB\log_2(1+\text{SNR}_2 z_2)}\right\}\right\} \quad (106)$$

$$= -\frac{1}{\theta_1}\log \mathbb{E}_{z_1}\left\{e^{-\tilde{\tau}\theta_1 TB\log_2(1+\text{SNR}_1 z_1)}\right\} \quad (107)$$

where $\tilde{\tau} = \min\{\tau_0, \tau^*\}$ and $\tau^*$ is the solution to

$$-\frac{1}{\theta_1}\log \mathbb{E}_{z_1}\left\{e^{-\tau\theta_1 TB\log_2(1+\text{SNR}_1 z_1)}\right\} = -\frac{1}{\theta_2}\log \mathbb{E}_{z_2}\left\{e^{-(1-\tau)\theta_2 TB\log_2(1+\text{SNR}_2 z_2)}\right\} \quad (108)$$

and $\tau_0$, as defined in (46), is the upper bound on the time-sharing parameter $\tau$. Above, (106) can be easily obtained by using a similar approach as in the proof of Proposition 1. (107) follows from the fact that the first term inside the minimization in (106) is an increasing function of $\tau$ while the second term is a decreasing function. Hence, the upper bound in (106) is maximized at $\tau^*$ at which the two terms inside the minimization are equal to each other. If $\tau^* < \tau_0$, the optimal value of $\tau$ is selected as $\tau^*$. If, on the other hand, $\tau^*$ exceeds the upper bound, i.e., $\tau^* \geq \tau_0$, then the optimal value is $\tau_0$.

**Case I** $\theta_1 \geq \theta_2$:

In this case in which the QoS constraint at the source is more stringent, we can show that the upper bound in (107) can be achieved or be approached arbitrarily closely. Let us set $\tilde{\theta} = \theta_1$, $\hat{\theta} = \theta_2$, and choose the time-sharing parameter as $\tau = \tilde{\tau} = \min\{\tau_0, \tau^*\}$. Now, the equation in (44) becomes

$$R = g(\theta_1) = -\frac{1}{\theta_1}\log \mathbb{E}_{z_1}\left\{e^{-\tilde{\tau}\theta_1 TB\log_2(1+\text{SNR}_1 z_1)}\right\}. \quad (109)$$

Since $\hat{\theta} = \theta_2 \leq \tilde{\theta} = \theta_1$ by our assumption in Case I, (45) reduces to

$$R = h(\theta_1, \theta_2) = -\frac{1}{\theta_2}\log \mathbb{E}_{z_2}\left\{e^{-(1-\tilde{\tau})\theta_2 TB\log_2(1+\text{SNR}_2 z_2)}\right\}. \quad (110)$$

Now, first assume that $\tilde{\tau} = \tau^*$. As seen in (108), we have, by the definition of $\tau^*$, that the right-hand sides of (109) and (110) are equal and therefore these equations are simultaneously satisfied.

Next, consider the other possibility in which $\tilde{\tau} = \min\{\tau_0, \tau^*\} = \tau_0$ which implies that $\tau_0 \leq \tau^*$. Note again that $\tau^*$ is the value of $\tau$ at which the functions

$$-\frac{1}{\theta_1} \log \mathbb{E}_{z_1} \left\{ e^{-\tau \theta_1 TB \log_2(1+\text{SNR}_1 z_1)} \right\} \tag{111}$$

and

$$-\frac{1}{\theta_2} \log \mathbb{E}_{z_2} \left\{ e^{-(1-\tau)\theta_2 TB \log_2(1+\text{SNR}_2 z_2)} \right\} \tag{112}$$

are equal. Note that the function in (111) increases with increasing $\tau$ while the function in (112) decreases. They meet at $\tau^*$. Therefore, at $\tau = \tau_0 \leq \tau^*$, we have

$$-\frac{1}{\theta_1} \log \mathbb{E}_{z_1} \left\{ e^{-\tau_0 \theta_1 TB \log_2(1+\text{SNR}_1 z_1)} \right\} \leq -\frac{1}{\theta_2} \log \mathbb{E}_{z_2} \left\{ e^{-(1-\tau_0)\theta_2 TB \log_2(1+\text{SNR}_2 z_2)} \right\}. \tag{113}$$

Hence, the rate

$$R = -\frac{1}{\theta_1} \log \mathbb{E}_{z_1} \left\{ e^{-\tau_0 \theta_1 TB \log_2(1+\text{SNR}_1 z_1)} \right\} \tag{114}$$

can be supported. More specifically, the equations in (44) and (45) can simultaneously be satisfied by setting $\tilde{\theta} = \theta_1$, $\tau = \tau_0$, and also by choosing $\hat{\theta} > \theta_2$ so that the right-hand side of (45) becomes smaller than $-\frac{1}{\theta_2} \log \mathbb{E}_{z_2} \left\{ e^{-(1-\tau_0)\theta_2 TB \log_2(1+\text{SNR}_2 z_2)} \right\}$ and matches $-\frac{1}{\theta_1} \log \mathbb{E}_{z_1} \left\{ e^{-\tau_0 \theta_1 TB \log_2(1+\text{SNR}_1 z_1)} \right\}$.

One subtlety in the above argument is the following. Note that we have the strict inequality $\tau < \tau_0$. Hence, we cannot actually set $\tau = \tau_0$ but we can select a value of $\tau$ that is arbitrarily close to $\tau_0$. Therefore, since the function in (111) increases with increasing $\tau$, we can approach the maximum rate $-\frac{1}{\theta_1} \log \mathbb{E}_{z_1} \left\{ e^{-\tau_0 \theta_1 TB \log_2(1+\text{SNR}_1 z_1)} \right\}$ arbitrarily closely. Because the effective capacity is defined as the supremum of rates (see e.g., (15)), $R = -\frac{1}{\theta_1} \log \mathbb{E}_{z_1} \left\{ e^{-\tau_0 \theta_1 TB \log_2(1+\text{SNR}_1 z_1)} \right\}$ is indeed the effective capacity.

**Case II** $\theta_1 < \theta_2$:

We now consider the scenario in which the relay node is subject to a more stringent QoS constraint. In this case, the approach behind the proof is identical to the one employed in Case I. Again, we set $\tilde{\theta} = \theta_1$ and $\hat{\theta} = \theta_2$. Because, otherwise if we have $\tilde{\theta} > \theta_1$ and/or $\hat{\theta} > \theta_2$, we impose more strict QoS constraints than necessary and hence end up supporting only lower arrival rates. Now, for fixed $\tau$, the equations in (44) and (45) become

$$R = g(\theta_1) = -\frac{1}{\theta_1} \log \mathbb{E}_{z_1}\left\{e^{-\tau\theta_1 TB \log_2(1+\text{SNR}_1 z_1)}\right\} \tag{115}$$

and

$$R = h(\theta_1, \theta_2) = -\frac{1}{\theta_1}\left(\log \mathbb{E}_{z_2}\left\{e^{-(1-\tau)\theta_2 TB \log_2(1+\text{SNR}_2 z_2)}\right\} + \log \mathbb{E}_{z_1}\left\{e^{\tau(\theta_2-\theta_1)TB \log_2(1+\text{SNR}_1 z_1)}\right\}\right), \tag{116}$$

respectively. Note that (116) follows from (45) by noting that $\hat{\theta} = \theta_2 > \theta_1 = \tilde{\theta}$ in this case. Similarly as before, the right-hand side of (115) is an increasing function of $\tau$ while the right-hand side of (116) is a decreasing function. Therefore, the equations in (115) and (116) can simultaneously be satisfied by choosing $\tau = \tau'$ where $\tau'$ is solution to

$$-\frac{1}{\theta_1} \log \mathbb{E}_{z_1}\left\{e^{-\tau\theta_1 TB \log_2(1+\text{SNR}_1 z_1)}\right\}$$
$$= -\frac{1}{\theta_1}\left(\log \mathbb{E}_{z_2}\left\{e^{-(1-\tau)\theta_2 TB \log_2(1+\text{SNR}_2 z_2)}\right\} + \log \mathbb{E}_{z_1}\left\{e^{\tau(\theta_2-\theta_1)TB \log_2(1+\text{SNR}_1 z_1)}\right\}\right). \tag{117}$$

Choosing values other than $\tilde{\theta} = \theta_1$, $\hat{\theta} = \theta_2$, and $\tau = \tau'$ will lead to smaller arrival rates. Hence, the effective capacity is given by

$$R_E(\theta_1, \theta_2) = -\frac{1}{\theta_1} \log \mathbb{E}_{z_1}\left\{e^{-\tau'\theta_1 TB \log_2(1+\text{SNR}_1 z_1)}\right\}. \tag{118}$$

Above discussion implicitly assumes that $\tau' < \tau_0$. If $\tau'$ exceeds the threshold $\tau_0$, then the optimal value of the time-sharing parameter is set to $\tau = \tau_0$. Using similar ideas as in Case I, we can show that the effective capacity in this case is

$$R_E(\theta_1, \theta_2) = -\frac{1}{\theta_1} \log \mathbb{E}_{z_1}\left\{e^{-\tau_0\theta_1 TB \log_2(1+\text{SNR}_1 z_1)}\right\}. \tag{119}$$

# REFERENCES

[1] A. Goldsmith, *Wireless Communications*, 1st ed. Cambridge University Press, 2005.

[2] C.-S. Chang, "Stability, queue length, and delay of deterministic and stochastic queuing networks," *IEEE Trans. Auto. Control*, vol. 39, no. 5, pp. 913-931, May 1994.

[3] C.-S. Chang and T. Zajic, "Effective bandwidths of departure processes from queues with time varying capacities," In *Proceedings of IEEE Infocom,* pp. 1001-1009, 1995.

[4] C.-S. Chang and J.A. Thomas, "Effective bandwidth in high-speed digital networks, "*IEEE Journal on Selected Areas in Communiations,* vol. 13, no. 6, Aug. 1995, pp. 1091-1100.

[5] D. Wu and R. Negi "Effective capacity: a wireless link model for support of quality of service," *IEEE Trans. Wireless Commun.*, vol.2,no. 4, pp.630-643. July 2003

[6] J. Tang and X. Zhang, "Quality-of-service driven power and rate adaptation over wireless links," *IEEE Trans. Wireless Commun.*, vol. 6, no. 8, pp. 3058-3068, Aug. 2007.

[7] J. Tang and X. Zhang, "Cross-layer-model based adaptive resource allocation for statistical QoS guarantees in mobile wireless networks," *IEEE Trans. Wireless Commun.*, vol. 7, no. 6, pp.2318-2328, June 2008.

[8] Q. Du and X. Zhang, "Resource allocation for downlink statistical multiuser QoS provisionings in cellular wireless networks," in Proc. IEEE International Conference on Computer Communications (INFOCOM) 2008, Phoenix, Arizona, USA, Apr. 13 -19, 2008, pp. 2405-2413.

[9] M.C. Gursoy, D. Qiao, and S. Velipasalar, "Analysis of energy efficiency in fading channel under QoS constrains," *IEEE Trans. on Wireless Commun.*, vol. 8, no. 8, pp. 4252-4263, Aug. 2009.

[10] E. A. Jorswieck, R. Mochaourab, and M. Mittelbach, "Effective capacity maximization in multi-antenna channels with covariance matrix, "*IEEE Trans. Wireless Commun.*, vol. 9, no. 10, pp. 2988-2993, Oct. 2010.

[11] A. Balasubramanian and S.L. Miller, "The effective capacity of a time division downlink scheduling system, "*IEEE Trans. Commun.*, vol. 58, no. 1, pp. 73-78, Jan. 2010.

[12] J. Tang and X. Zhang, "Cross-layer resource allocation over wireless relay networks for quality of service provisioning, "*IEEE Journal on Selected Areas in Communications*, vol. 25, no. 4, pp. 645-656, May 2007.

[13] P. Parag, and J.-F. Chamberland, "Queueing analysis of a butterfly network for comparing network coding to classical routing," *IEEE Trans. Inform. Theory*, vol. 56, no. 4, pp. 1890-1907, Apr. 2010.